\documentclass[proof]{pasj01}
\usepackage{url}
\usepackage{multirow}
\usepackage{lscape}
\usepackage{ulem}
\usepackage{natbib}
\usepackage{subfigure}
\usepackage[figuresright]{rotating}
\Received{$\langle$reception date$\rangle$}
\Accepted{$\langle$acception date$\rangle$}
\Published{$\langle$publication date$\rangle$}
\begin{document}

\title{New timing measurement results of 16 pulsars}
\author{Jie Liu$^{1,2,3}$, Zhen Yan$^{1,2,7 *}$\email{yanzhen@shao.ac.cn}, Zhi-Qiang Shen$^{1,2,3,7}$, Zhi-Peng Huang$^{1,2,3}$, Ru-Shuang Zhao$^{6}$, Ya-Jun Wu$^{1}$, Jian-Ping Yuan$^{4,7}$, Xin-Ji Wu$^{5}$} %
\altaffiltext{1}{Shanghai Astronomical Observatory, Chinese Academy of Sciences, Shanghai, China}
\altaffiltext{2}{University of Chinese Academy of Sciences, Beijing, China}
\altaffiltext{3}{School of Physical Science and Technology, ShanghaiTech University, Shanghai, China}
\altaffiltext{4}{Xinjiang Astronomical Observatory, Chinese Academy of Sciences, Urumqi, China}
\altaffiltext{5}{School of Physics, Peking University, Beijing, China}
\altaffiltext{6}{School of Physics and Electronic Science, Guizhou Normal University, Guiyang, China}
\altaffiltext{7}{Key Laboratory of Radio Astronomy, Chinese Academy of Sciences, Nanjing, China}

\maketitle

\begin{abstract}
Pulsar's position, proper motion and parallax are important parameters in timing equations. It is a really challenging work to fit astrometric parameters accurately through pulsar timing, especially for pulsars that show irregular timing properties. As the fast development of related techniques, it is possible to measure astrometric parameters of more and more pulsars in a model$\textrm{-}$independent manner with the Very Long Baseline Interferometry (VLBI). In this work, we select 16 normal pulsars, whose parallax and proper motion have not been successfully fitted with timing observations or show obvious differences with corresponding latest VLBI solutions, and do further studies on their timing properties. After updating astrometric parameters in pulsar ephemerides with the latest VLBI measurements, we derive the latest rotation solutions of these pulsars with observation data at S and C$\textrm{-}$band obtained from the Shanghai Tian Ma Radio Telescope (TMRT). Compared with spin frequency $\nu$ inferred from previous rotation solutions, the newly$\textrm{-}$fitted $\nu$ show differences larger than 10$^{-9}$~Hz for most pulsars. The contribution of the Shklovsky effect to period derivative $\dot{P}$ can be properly removed taking advantages of accurate proper motion and distance of target pulsars measured by VLBI astrometry. This further leads to a precise estimate of intrinsic characteristic age $\tau_{\rm c}$. Differences between the newly$\textrm{-}$measured $\tau_{\rm c}$ and corresponding previous results are as large as 2\% for some pulsars. VLBI astrometric parameter solutions also lead to better measurements of timing irregularities. For PSR B0154$+$61, the glitch epoch (MJD 58279.5) measured with previous ephemeris is about 13~d later than the result (MJD 58266.4) obtained with VLBI astrometric parameter solutions.
\end{abstract}

\section{Introduction}

Pulsar timing is the process of measuring pulse time of arrivals (ToAs) and identifying the phenomena that affect them. It is widely used in measuring the mass of solar system planets \citep{chm+10}, detecting gravitational waves \citep{det79}, constructing pulsar$\textrm{-}$based time$\textrm{-}$scale \citep{hcm+12} and so on. For solitary normal pulsar, ToAs at the solar system barycenter ($t_{\rm ssb}$) are transformed from observed ToAs ($t_{\rm o}$) by
\begin{equation}
t_{\rm ssb}= t_{\rm o}+t_{\rm c}-\Delta D/f^{2}+\Delta_{\rm R \odot}(\alpha,\delta,\mu_{\alpha},\mu_{\delta},\pi)  +\Delta_{\rm S \odot}(\alpha,\delta)+\Delta_{\rm E \odot}
\label{eq:equation 1}
\end{equation}
where, $t_{\rm c}$, $\Delta D/f^{2}$, $\Delta_{\rm R \odot}$, $\Delta_{\rm S \odot}$ and $\Delta_{\rm E \odot}$ are clock correction due to variations of local hydrogen atomic clock, correction due to dispersion in the interstellar medium, the time that light travels from the phase center of receiver to the solar system barycenter (R$\ddot{\rm o}$mer delay), time delay due to the space$\textrm{-}$time curvature caused by massive objects in the solar system (Shapiro delay) and combined time correction due to earth motion and gravitational redshift caused by other solar system objects (Einstein delay), respectively. $\Delta_{\rm R \odot}$ and $\Delta_{\rm S \odot}$ are functions of astrometric parameters including position ($\alpha$, $\delta$), proper motion ($\mu_{\alpha}$, $\mu_{\delta}$) and parallax $\pi$ of the target pulsar. For a binary pulsar, this equation should be extended with additional terms caused by companion and its orbital motion \citep{tay89}.

It is challenging to obtain accurate pulsar astrometric parameters. Usually, pulsar distance is roughly estimated based on its dispersion measure (DM) and Galactic electron density distribution models, such as TC93 \citep{tc93}, NE2001 \citep{cl02} and YMW16 \citep{ymw17}. However, precision of distance measured by model$\textrm{-}$dependent methods is limited by low fidelity of Galactic electron density distribution models. What's more, DM$\textrm{-}$based distance of some individual pulsars may have great error \citep{dtbr09} or systematic bias \citep{lfl+06}. The lower or upper limits of distance can also be estimated by H\uppercase\expandafter{\romannumeral1} absorption measurements for some pulsars \citep{kjww95}. Pulsar's motion can be estimated by measuring interstellar scintillation effect \citep{cw84,cor86,gg00,ob05}. However, scintillation velocity has large error and highly depends on measurement of pulsar's distance \citep{gup95}.

Except the optical method for few pulsars (or their companion stars) with corresponding radiation, there are two common ways to measure astrometric parameters of pulsars. One is long$\textrm{-}$term high$\textrm{-}$accuracy pulsar timing, the other is multi$\textrm{-}$epoch high$\textrm{-}$resolution VLBI observations. As what is shown in Equation~\ref{eq:equation 1}, pulsar astrometric parameters are not independent of each other. It is difficult to precisely measure pulsar astrometric parameters by pulsar timing, especially for pulsars that show timing irregularities (timing noise or glitches). The position and proper motion of 374 pulsars were measured using extremely long$\textrm{-}$term timing data obtained from the 76$\textrm{-}$m Lovell Radio Telescope at Jodrell Bank Observatory \citep{hlk+04}. This is the largest sample of pulsar astrometric parameters obtained with pulsar timing so far. Although the ``harmonic whitening" method was used in their data analysis process, fitted astrometric parameters of some pulsars still had large errors. It was almost impossible to obtain relatively reliable proper motion for some of these pulsars.

High$\textrm{-}$precision VLBI astrometry offers a powerful way to directly measure the parallax and proper motion of a pulsar. The efforts of measuring pulsar astrometric parameters by radio interferometry started in 1970s \citep{alp75,bs76}. Initially, it was only possible to perform interferometry observations of several bright nearby pulsars. With steady progress of VLBI observations, correlation and data processing techniques \citep{brs+85, dtbw07}, astrometry observations of more and more pulsars have been successfully accomplished with the VLBI \citep{gtwr86,slh+99,bfg+03,cbv+09,dtbr09,dgb+19}. The precision of VLBI measurements is getting higher and higher. By now, there are at least 96 pulsars whose astrometric parameters have been accurately measured with the VLBI \footnote{http://hosting.astro.cornell.edu/$\sim$shami/psrvlb/parallax.html}. Benefiting from the high accuracy and independence of pulsar astrometric parameters obtained with the VLBI, PSR J0337$+$1715 was successfully identified as a millisecond pulsar stellar triple system by pulsar timing \citep{rsa+14}.

It is expectable that timing properties of normal pulsars will be better estimated by updating their astrometric parameters using VLBI measurements, as this has been proved useful for millisecond pulsars. We select 16 normal pulsars whose parallax and proper motion have not been successfully fitted with timing observations or differ a lot from corresponding latest VLBI results \citep{hlk+04, dgb+19}. The pulsar name, timing and VLBI solutions of position (R.A., Dec), proper motion ($\mu_{\rm \alpha}$, $\mu_{\rm \delta}$), parallax (PX) and position epoch (Pos epoch) are listed in Table~\ref{Tab:table 1} from left to right column, respectively. These 16 pulsars are frequently observed with the newly built Shanghai Tian Ma Radio Telescope (TMRT). We update position, proper motion and parallax in ephemerides (referred as updated ephemerides) of 16 pulsars using their latest VLBI measurements, and do further studies on their timing properties. This paper is organized in the structure as follows. Observations and data reduction are introduced in Section~\ref{sect:Observation}. Analysis and detailed results are shown in Section~\ref{sect:results}. Discussion and conclusion are presented in Sections~\ref{sect:discussion} and~\ref{sect:conclusion}, respectively.

\begin{table*}[htb]
\footnotesize{
\caption{Timing and the latest VLBI astrometric parameter solutions of 16 pulsars.}
\centering
\begin{tabular}{c c l l l l l c}
\hline
\hline
Name & Solutions & \multicolumn{1}{c}{R.A.} & \multicolumn{1}{c}{Dec} & \multicolumn{1}{c}{$\mu_{\alpha}$} & \multicolumn{1}{c}{$\mu_{\delta}$} & \multicolumn{1}{c}{PX} & \multicolumn{1}{c}{Pos epoch}\\
 & & \multicolumn{1}{c}{(h~m~s)} & \multicolumn{1}{c}{(d~m~s)} & \multicolumn{1}{c}{(mas~yr$^{-1}$)} & \multicolumn{1}{c}{(mas~yr$^{-1}$)}  & \multicolumn{1}{c}{(mas)} & \multicolumn{1}{c}{(MJD)}\\
\hline
\multirow{2}{*}{B0031$-$07} & Timing & 00:34:08.84(5) & $-$07:21:53.1(14) & $-$16(26)    & 17(53) & * & \multirow{2}{*}{52275} \\
 & VLBI$^{\rm b}$ & 00:34:08.8703(1) & $-$07:21:53.409(2) & 10.37(8)    & $-$11.13(16) & 0.93(8) & \\
\hline
\multirow{2}{*}{B0136$+$57} & Timing & 01:39:19.744(4)& $+$58:14:31.73(3)& * & *   & * & \multirow{2}{*}{52275} \\
 & VLBI$^{\rm b}$ & 01:39:19.7401(12)& $+$58:14:31.819(17)& $-$19.11(7) &$-$16.60(7)   & 0.37(4) &  \\
\hline
\multirow{2}{*}{B0148$-$06} & Timing & 01:51:22.72(6) & $-$06:35:03.5(20) & 19(42)     & $-$39(87)   & * &\multirow{2}{*}{56000} \\
 & VLBI$^{\rm c}$ & 01:51:22.7179(2) & $-$06:35:02.987(2) & 10.7(1)     & $-$5.38(7)   & 0.23(9) &  \\
\hline
\multirow{2}{*}{B0154$+$61} & Timing  & 01:57:49.80(7) & $+$62:12:27.3(7) & $-$51(36)     & 81(33)     & * & \multirow{2}{*}{56000}\\
 & VLBI$^{\rm c}$ & 01:57:49.9434(1) & $+$62:12:26.648(1) & 1.57(6)     & 44.80(4)     & 0.56(3) &  \\
\hline
\multirow{2}{*}{B0450$+$55} & Timing & 04:54:07.746(6) & $+$55:43:41.4(1) & 48(6)    & $-$13(12) & * &  \multirow{2}{*}{52275} \\
 & VLBI$^{\rm b}$ & 04:54:07.7506(1) & $+$55:43:41.437(2) & 53.34(6)    & $-$17.56(14) & 0.84(5) & \\
\hline
\multirow{2}{*}{B0611$+$22} & Timing & 06:14:17.16(3) & $+$22:30:36(17) & *  & *   & * & \multirow{2}{*}{56000} \\
 & VLBI$^{\rm c}$ & 06:14:17.0058(1) & $+$22:29:56.848(1) & $-$0.24(4)  & $-$1.25(4)   & 0.28(3) &  \\
\hline
\multirow{2}{*}{B0626$+$24} & Timing & 06:29:05.77(2) & $+$24:15:50(6) & 32(15)    & 400(300)   & * & \multirow{2}{*}{56000}\\
 & VLBI$^{\rm c}$ & 06:29:05.7273(1) & $+$24:15:41.546(1) & 3.56(12)    & $-$4.68(8)   & 0.32(5) &  \\
\hline
\multirow{2}{*}{B0727$-$18} & Timing & 07:29:32.351(3) & $-$18:36:42.75(7) & * & *   & * & \multirow{2}{*}{56000} \\
 & VLBI$^{\rm c}$ & 07:29:32.3369(1) & $-$18:36:42.244(2) & $-$13.06(11)& 13.27(44)    & 0.50(9) &  \\
\hline
\multirow{2}{*}{B0809$+$74} & Timing & 08:14:59.49(2)& $+$74:29:05.8(2)& 18(20)    & $-$49(21)   & * &  \multirow{2}{*}{48382} \\
 & VLBI$^{\rm a}$ & 08:14:59.5412(10)& $+$74:29:05.367(15)& 24.02(9)    & $-$44.0(4)   & 2.31(4) & \\
\hline
\multirow{2}{*}{B0820$+$02} & Timing & 08:23:09.77(1) & $+$01:59:12.7(5) & $-$2(8)   & 18(24)  & *  & \multirow{2}{*}{56000}  \\
 & VLBI$^{\rm c}$ & 08:23:09.7651(1) & $+$01:59:12.469(1) & $-$4.0(2)   & 0.15(26)     & 0.4(1)  &  \\
\hline
\multirow{2}{*}{B1530$+$27} & Timing & 15:32:10.37(4) & $+$27:45:49.3(7) & 6(24)      & $-$7(30)    & * & \multirow{2}{*}{56000} \\
 & VLBI$^{\rm c}$ & 15:32:10.3646(1) & $+$27:45:49.623(1) & 1.5(1)      & 18.93(11)    & 0.59(6) &  \\
\hline
\multirow{2}{*}{B1540$-$06} & Timing & 15:43:30.136(9) & $-$06:20:45.5(4) & $-$18(6) & $-$14(19)  & * &  \multirow{2}{*}{56000} \\
 & VLBI$^{\rm c}$ & 15:43:30.1373(1) & $-$06:20:45.332(2) & $-$16.79(4) & $-$0.30(13)  & 0.31(4) & \\
\hline
\multirow{2}{*}{B1541$+$09} & Timing & 15:43:38.82(4) & $+$09:29:15(1) & 9(39)  & $-$113(76)   & * & \multirow{2}{*}{52275}\\
 & VLBI$^{\rm b}$ & 15:43:38.8250(1) & $+$09:29:16.339(2) & $-$7.61(6)  & $-$2.87(7)   & 0.13(2) &  \\
\hline
\multirow{2}{*}{B1642$-$03} & Timing & 16:45:02.0414(6) & $-$03:17:58.32(3) & * & *     & * &  \multirow{2}{*}{56000} \\
 & VLBI$^{\rm c}$ & 16:45:02.0406(1) & $-$03:17:57.819(2) & $-$1.035(27)& 20.5(2)      & 0.26(2) &  \\
\hline
\multirow{2}{*}{B2154$+$40} & Timing & 21:57:01.84(1) & $+$40:17:45.9(2) & 0(13)   & 7(15)     & * & \multirow{2}{*}{52275} \\
 & VLBI$^{\rm b}$ & 21:57:01.8495(1) & $+$40:17:45.986(2) & 16.13(10)   & 4.12(12)     & 0.28(6) &   \\
\hline
\multirow{2}{*}{B2351$+$61} & Timing & 23:54:04.77(1) & $+$61:55:46.8(1) & 18(8)    & $-$1(6)    & * & \multirow{2}{*}{56000} \\
 & VLBI$^{\rm c}$ & 23:54:04.7830(1) & $+$61:55:46.845(1) & 22.76(5)    & 4.897(25)    & 0.41(4) &  \\
\hline
\end{tabular}
\label{Tab:table 1}
\\
Reference: a. \cite{bbgt02}; b. \cite{cbv+09}; c. \cite{dgb+19}. Timing solutions are quoted from \cite{hlk+04}. The symbol * means that parameters are not provided. Timing positions are inferred to same epoch of corresponding VLBI solutions for each pulsar.
}
\end{table*}

\section{Observations and Data reduction}
\label{sect:Observation}

Pulsar timing observations were carried out at S and C$\textrm{-}$band with the 65$\textrm{-}$m TMRT between MJD 57292 (2015 September 27) and 58866 (2020 January 18). Both S and C$\textrm{-}$band receivers are cryogenically cooled dual$\textrm{-}$polarization receivers. The effective frequency coverages of S$\textrm{-}$band and C$\textrm{-}$band receivers are 2200$\textrm{-}$2300~MHz and 4320$\textrm{-}$5320~MHz, respectively. For the convenience in performing online dedispersion and removing radio frequency interference (RFI), the full bandwidths of S and C$\textrm{-}$band receivers were correspondingly divided into channels with the typical width of 1 and 2~MHz in the Digital Backend System (DIBAS). Data sampling and recording were accomplished with the DIBAS. For both S and C$\textrm{-}$band pulsar observations, the time resolution of data sampling was 40.96~us \citep{ysw+18}. Data were obtained under the incoherent dedispersion on$\textrm{-}$line folding observation mode with the sub$\textrm{-}$integration time of 30~s. The observational files were written out as 8$\textrm{-}$bit {\sc PSRFITS} format with 1024 phase bins for each period \citep{ysw+15}. Pulsar parameters used in online folding observations were obtained from the {\sc PSRCAT}\footnote{http://www.atnf.csiro.au/research/pulsar/psrcat/} \citep{mhth05}. Depending on the flux density of target pulsars and observation conditions (weather and RFI), observation durations were mostly ranging from 10 to 20~min.

During pulsar observations, time was marked using a local hydrogen atomic clock corrected to GPS. The {\sc PSRCHIVE} \citep{hvm04} and {\sc Tempo2} \citep{hob12} packages were used for data reduction and analysis. For each pulsar, off$\textrm{-}$line data from all channels and subintegrations were summed to produce mean pulse profile for each observation. Local ToAs were determined by the cross$\textrm{-}$correlation of observed pulse profiles with pulse template and were converted to ToAs at the solar system barycenter with the Jet Propulsion Laboratory's DE421 ephemeris.

\section{Analysis and Results}
\label{sect:results}
Timing analysis of 16 pulsars was performed using S and C$\textrm{-}$band timing data obtained from the TMRT. Pulsar ephemerides were quoted from the {\sc PSRCAT} (Version: 1.59). Astrometric parameters in ephemerides were updated using VLBI solutions in Table~\ref{Tab:table 1}. The ephemeris after updating astrometric parameters is referred as updated ephemeris. For pulsars that have both S and C$\textrm{-}$band data, timing residuals showed phase offset between data at different frequencies. The offset was fitted at the first step and kept constant in ephemeris with measured value. For all pulsars, $\nu$ and $\dot{\nu}$ were fitted based on updated ephemerides. The epoch of rotation parameters was set to be the middle date of data$\textrm{-}$span. According to their different timing properties, these 16 pulsars were classified into three groups: pulsar with glitch, pulsar with loud timing noise and pulsars without significant timing irregularities. Proper methods were used in the data analysis of pulsars depending on their timing properties.

\begin{figure}[h]
\begin{center}
\begin{tabular}{c}
\resizebox{0.9\hsize}{!}{\includegraphics[angle=0]{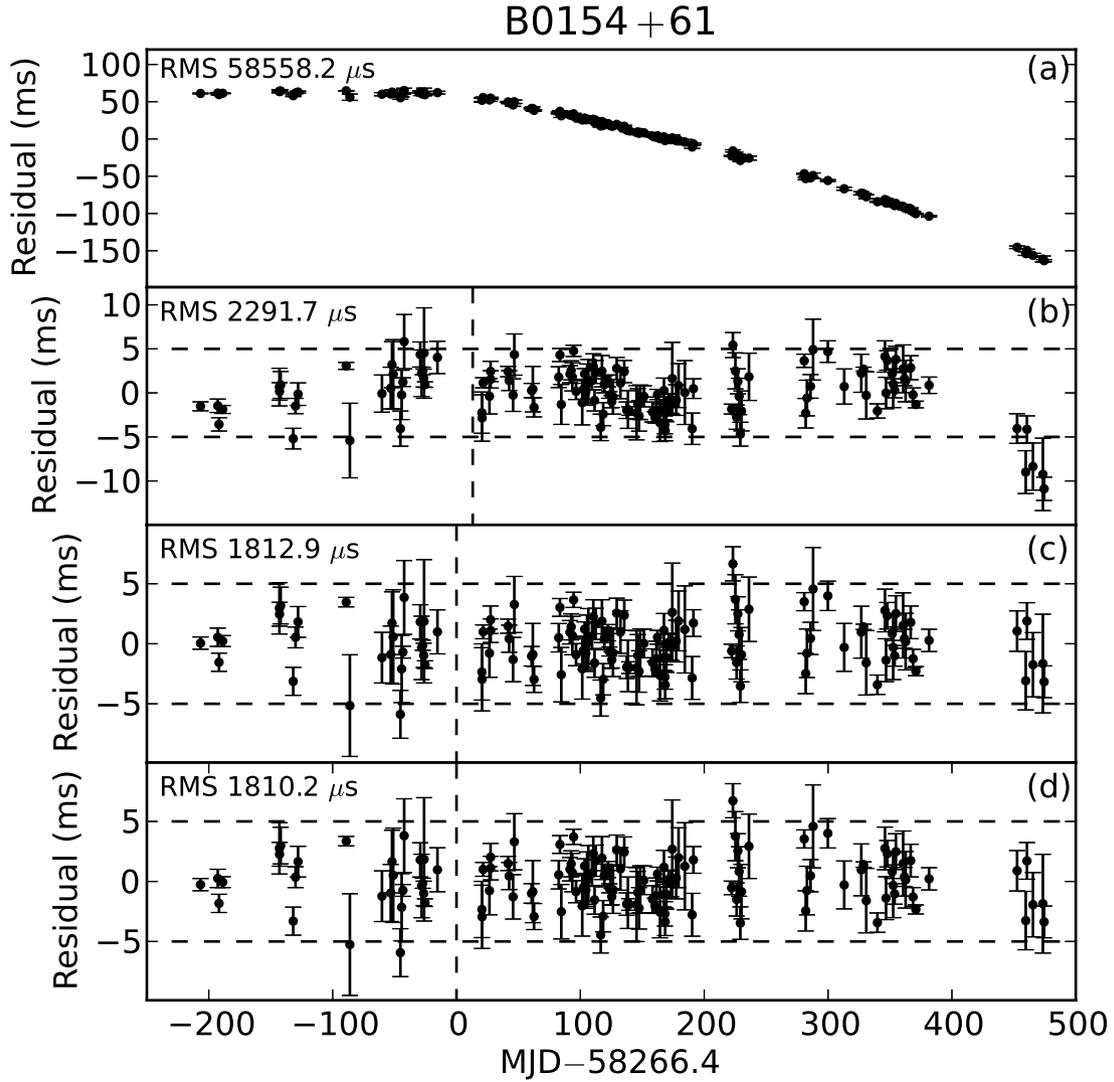}} \\
\end{tabular}
\end{center}
\caption{Timing residuals of PSR B0154$+$61 based on previous timing ephemeris without fitting glitch (a) and with glitch fitted (b). Panel (c) shows residuals based on updated ephemeris with glitch fitted. Panel (d) are residuals in (c) with $\nu$ and $\dot{\nu}$ further fitted. Dashed vertical lines indicate measured epoch of glitch. Horizontal dashed lines are plotted for comparing residuals and are at $\pm$5~ms, respectively. }
\label{fg:figure 1}
\end{figure}

(1) Pulsar with glitch\\
One glitch occurred in PSR B0154$+$61 around MJD 58266.4. Because of incorrect rotation parameters in updated ephemeris, timing residuals of PSR B0154$+$61 showed an obvious downward trend, in which, the glitch was hard to see. Rotation parameters of PSR B0154$+$61 were firstly fitted using pre$\textrm{-}$glitch timing data and kept constant in ephemeris with newly fitted values. Then, glitch parameters were fitted based on this new ephemeris. Measured glitch parameters were kept constant in ephemeris when obtaining final rotation solution. In order to study the influence of different ephemerides on glitch fitting, we also fitted glitch based on timing ephemeris, in which, astrometric parameters are timing solutions in Table~\ref{Tab:table 1}. Rotation parameters in timing ephemeris are the same as rotation parameters in updated ephemeris. Table~\ref{Tab:table 2} shows glitch parameters measured with timing and updated ephemerides. The glitch epoch, $\Delta \nu$ and $\Delta \dot{\nu}$ are MJD 58266.4, 1.45$\times$10$^{-9}$~Hz and 4.1$\times$10$^{-17}$~s$^{-2}$ when measured with updated ephemeris, respectively. They are MJD 58279.5, 1.75$\times$10$^{-9}$~Hz and 2.6$\times$10$^{-17}$~s$^{-2}$ when measured with timing ephemeris, respectively. Figure~\ref{fg:figure 1} displays residuals before and after fitting glitch. Residuals in panel (a) show a sudden downward trend that caused by the glitch. In panel (b), residuals show noticeable fluctuations after fitting glitch with timing ephemeris. While, no obvious fluctuation is left based on updated ephemeris in panel (c). Residuals after further fitting $\nu$ and $\dot{\nu}$ in panel (d) have almost no difference with residuals in panel (c), implying that final rotation parameters are similar to that derived with pre$\textrm{-}$glitch data.

\begin{figure}[htb]
\begin{center}
\begin{tabular}{c}
\resizebox{0.9\hsize}{!}{\includegraphics[angle=0]{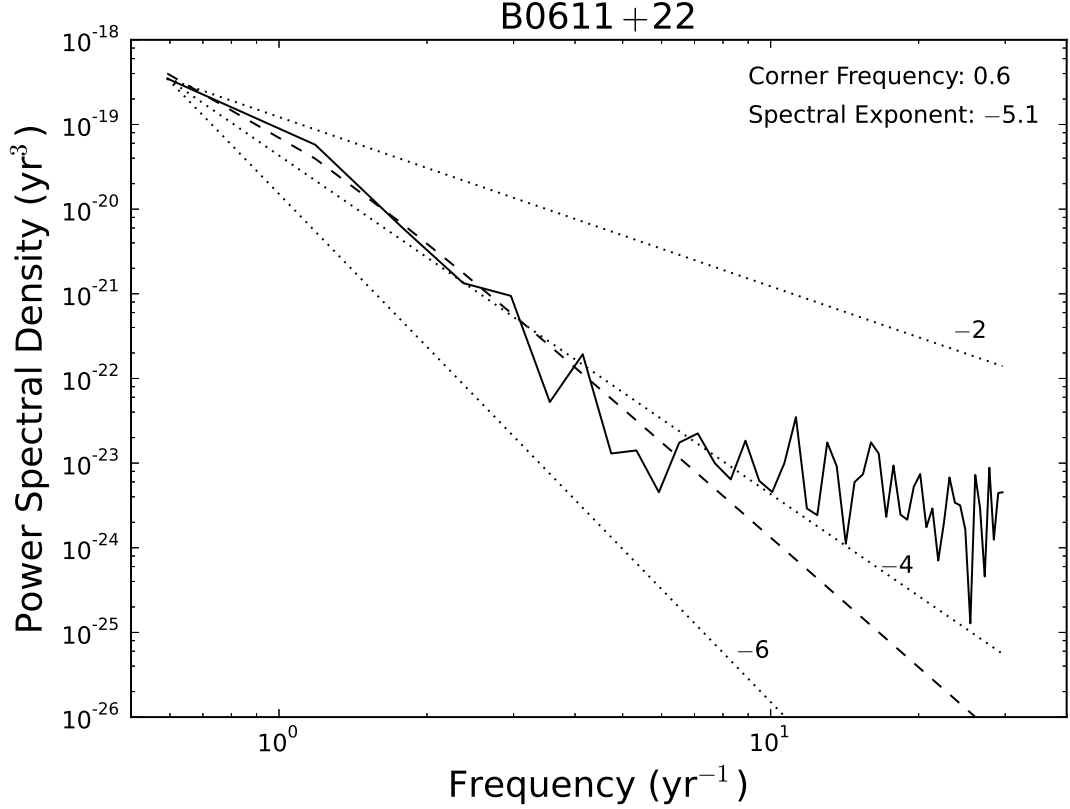}} \\
\end{tabular}
\end{center}
\caption{Observed and fitted power spectra of timing noise for PSR B0611$+$22. Solid jagged line represents spectrum of the smoothed and interpolated red noise. The power$\textrm{-}$law model fitted to red component is shown as dashed line. Dotted lines are power spectra with exponent of $-$2, $-$4 and $-$6, respectively.}
\label{fg:figure 2}
\end{figure}

\begin{figure}[ht]
\begin{center}
\begin{tabular}{c}
\resizebox{0.9\hsize}{!}{\includegraphics[angle=0]{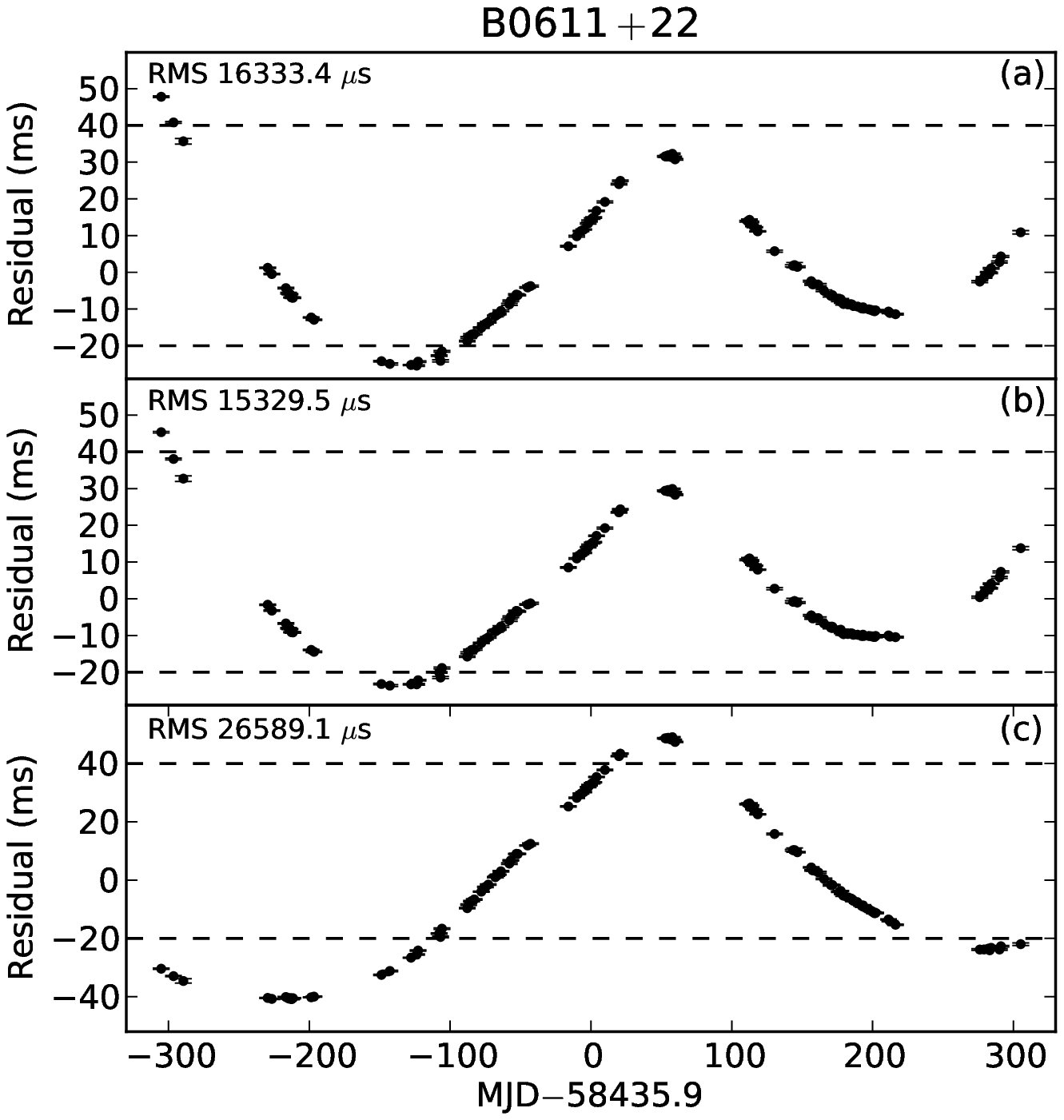}} \\
\end{tabular}
\end{center}
\caption{Panels (a) and (b) are timing residuals based on previous timing ephemeris and updated ephemeris without loading power$\textrm{-}$law spectrum model, respectively. Panel (c) shows residuals after fitting $\nu$ and $\dot{\nu}$ with spectrum model loaded. Horizontal dashed lines are at 40~ms and $-$20~ms, respectively. }
\label{fg:figure 3}
\end{figure}

(2)Pulsar with loud timing noise\\
PSR B0611$+$22 is a pulsar with loud timing noise. Due to incorrect rotation parameters in updated ephemeris, we firstly fitted the $\nu$ and $\dot{\nu}$ based on updated ephemeris. Fitted rotation parameters were kept constant in ephemeris. Timing noise can be partly absorbed into rotation parameters if these parameters are determined by standard pulsar timing procedures which neglect correlation in timing residuals. The Cholesky method provides the estimation of the covariance matrix of the timing noise, employs transformation of the matrix to whiten the timing residuals and improve the accuracy of timing parameters \citep{chc+11}. We used the Cholesky method to analyse timing noise. The spectrum of the red component of timing noise was fitted by a power$\textrm{-}$law model \citep{chc+11} 
\begin{equation}
P(f)=A/[1+(f/f_{\rm c})^{2}]^{\alpha /2}
\label{eq:equation 2}
\end{equation}
where, $A$, $f_{\rm c}$ and $\alpha$ are the amplitude, corner frequency and spectral exponent, respectively. The fit was accomplished using a {\sc Tempo2} plugin named {\sc SpectralModel}. Fitted power$\textrm{-}$law spectrum of PSR B0611$+$22 is shown in Figure~\ref{fg:figure 2}. The spectrum model of PSR B0611$+$22 was loaded using another plugin {\sc plk} when obtaining final rotation solution. We compared timing residuals based on timing ephemeris and updated ephemeris without loading the spectrum model. They are plotted in panels (a) and (b) of Figure~\ref{fg:figure 3}, respectively. The declination of PSR B0611$+$22 has a great difference of about 39$^{\prime\prime}$ between timing and VLBI solutions. Due to the existence of loud timing noise, differences between timing residuals based on timing and updated ephemerides are imperceptible compared with intrinsic timing noise (see panels (a) and (b)). However, RMS residuals in two panels have a large difference about 1~ms because of astrometric parameter differences between timing and VLBI solutions. In panel (c), structures of residuals after fitting $\nu$ and $\dot{\nu}$ with the spectrum model loaded show significant differences with that in panel (b). The RMS residual in panel (c) is about 11~ms lager than that in panel (b). This is because obvious trend may appear due to significant changes of rotation parameters after loading the spectrum model. Correspondingly, the amplitude of residual fluctuations will increase \citep{chc+11}. It is necessary to mention that the improvement on PSR B0611$+$22 is not remarkable after applying the Cholesky method due to short data span. It is expectable that the red noise will be more sufficiently characterized with more data accumulated for longer span in future.

(3) Pulsars without significant timing irregularities\\
Although PSRs B0136$+$57, B0450$+$55 and B1642$-$03 have obvious timing noise, their timing noise are much weaker compared with timing noise of PSR B0611$+$22 \citep{hlk10}. They were not fitted with power$\textrm{-}$law spectrum model. For pulsars without significant timing irregularities, we only fitted their rotation parameters using updated ephemerides. Measured rotation parameters were kept constant in both timing and updated ephemerides. We compared residuals based on timing ephemeris with corresponding residuals based on updated ephemeris for each pulsar. They are plotted as panels (a) and (b) in Figure~\ref{fg:figure 4}, respectively.

\textbf{PSR B0031$-$07:} Both $\mu_{\alpha}$ and $\mu_{\delta}$ measured by the VLBI have opposite directions to timing solutions. The declination difference between two solutions is about 0.3$^{\prime\prime}$. Differences between residuals shown in panels (a) and (b) of this pulsar are not obvious.

\textbf{PSR B0136$+$57:} The position difference between timing and VLBI solutions is not large. As timing proper motion is not provided, differences of $\mu_{\alpha}$ and $\mu_{\delta}$ are 19.11~mas~yr$^{-1}$ and 16.6~mas~yr$^{-1}$ between two solutions, respectively. The amplitude of residual fluctuations in panel (a) is larger than that of residual fluctuations in panel (b).

\textbf{PSR B0148$-$06:} For this pulsar, differences of declination, $\mu_{\alpha}$ and $\mu_{\delta}$ are about 0.5$^{\prime\prime}$, 8.3~mas~yr$^{-1}$ and 33.6~mas~yr$^{-1}$ between two solutions, respectively. Although the difference between RMS residuals in panels (a) and (b) is about 0.1~ms, it's hard to discern differences between residuals in two panels.

\textbf{PSR B0450$+$55:} Both position and proper motion differences between timing and VLBI solutions are small. There are no obvious differences between residuals in two panels.

\textbf{PSR B0626$+$24:} There are large position and proper motion differences between two solutions. Differences of declination and $\mu_{\delta}$ are about 8.5$^{\prime\prime}$ and 395~mas~yr$^{-1}$, respectively. However, timing residuals based on two kinds of ephemerides only differ a little.

\textbf{PSR B0727$-$18:} Differences of declination, $\mu_{\alpha}$ and $\mu_{\delta}$ are about 0.5$^{\prime\prime}$, 13~mas~yr$^{-1}$ and 13.3~mas~yr$^{-1}$ between two solutions, respectively. Differences between residuals in panel (a) and residuals in panel (b) can't be discerned clearly.

\textbf{PSR B0809$+$74:} Differences of position and proper motion between timing and VLBI solutions are not significant by comparison. Residuals based on timing ephemeris show small differences compared with residuals based on updated ephemeris.

\textbf{PSR B0820$+$02:} The $\mu_{\delta}$ difference between two solutions is about 18~mas~yr$^{-1}$, while differences of position and $\mu_{\alpha}$ between two solutions are small. Residuals in panels (a) and (b) show imperceptible differences.

\textbf{PSR B1530$+$27:} The $\mu_{\delta}$ shows opposite directions in pulsar timing and VLBI solutions. Differences of declination, $\mu_{\alpha}$ and $\mu_{\delta}$ are about 0.3$^{\prime\prime}$, 4.5~mas~yr$^{-1}$ and 12~mas~yr$^{-1}$, respectively. There are no obvious differences between residuals based on timing and updated ephemerides.

\textbf{PSR B1540$-$06:} Differences of position and $\mu_{\alpha}$ between timing and VLBI solutions are really small. Difference of $\mu_{\delta}$ between two solutions is about 14~mas~yr$^{-1}$. We can only find some imperceptible differences between residuals in two panels.

\textbf{PSR B1541$+$09:} Declination and $\mu_{\delta}$ differences between two solutions of this pulsar are about 1.3$^{\prime\prime}$ and 110~mas~yr$^{-1}$, respectively. There are sinusoidal fluctuations in residuals based on timing ephemeris, while residuals based on updated ephemeris have no obvious fluctuations.

\textbf{PSR B1642$-$03:} This pulsar has a declination difference of 0.5$^{\prime\prime}$ and $\mu_{\delta}$ difference of 20.5~mas~yr$^{-1}$ between two solutions. Residual fluctuations in panel (a) are obviously different from that in panel (b).

\textbf{PSR B2154$+$40:} Differences of astrometric parameters between timing and VLBI solutions are small except the $\mu_{\alpha}$, of which, the difference is 16.13~mas~yr$^{-1}$. It is difficult to see differences between residuals in panels (a) and (b).

\textbf{PSR B2351$+$61:} Timing position and proper motion of this pulsar differ a little from VLBI solutions. Only small differences can be seen between residuals in two panels.

\begin{table}[ht]
\footnotesize{
\caption{Glitch solutions of PSR B0154$+$61 based on two ephemerides.}
\label{Tab:table 2}
\centering
\begin{tabular}{l l l}
\hline
\hline
\multicolumn{1}{c}{Parameter} &
\multicolumn{1}{c}{Timing} &
\multicolumn{1}{c}{Updated} \\
 & ephemeris & ephemeris \\
\hline
Glitch epoch (MJD)                          & 58279.5(7)       & 58266.4(5) \\
$\Delta \nu$ ($10^{-9}$~Hz)                & 1.75(5)       & 1.45(2) \\
$\Delta \nu / \nu$ ($10^{-9}$)             & 4.11(11)       & 3.41(5) \\
$\Delta \dot{\nu}$ ($10^{-17}$~s$^{-2}$)   & 2.6(3)        & 4.1(1) \\
$\Delta \dot{\nu} / \dot{\nu}$ ($10^{-3}$) & $-$0.75(7)     & $-$1.19(4) \\
RMS residual ($\mu$s)                      & 2291.7        & 1812.9                                                                                                                                                                                                                                                                                                                                                                                                                                                                                                                                                                                                                                                                                                                                                                                                                                                                                                                                                                                                                                                                                                                                                                                                                                                                                                                                                                                                                                                                                                                                                                                                                                                                                                                                                                                                                                                                                                                                                                                 \\
\hline
\end{tabular}
\\
}
\end{table}

\begin{figure*}[]
\begin{center}
\subfigure{
\begin{tabular}{cc}
\resizebox{0.45\hsize}{!}{\includegraphics[angle=0]{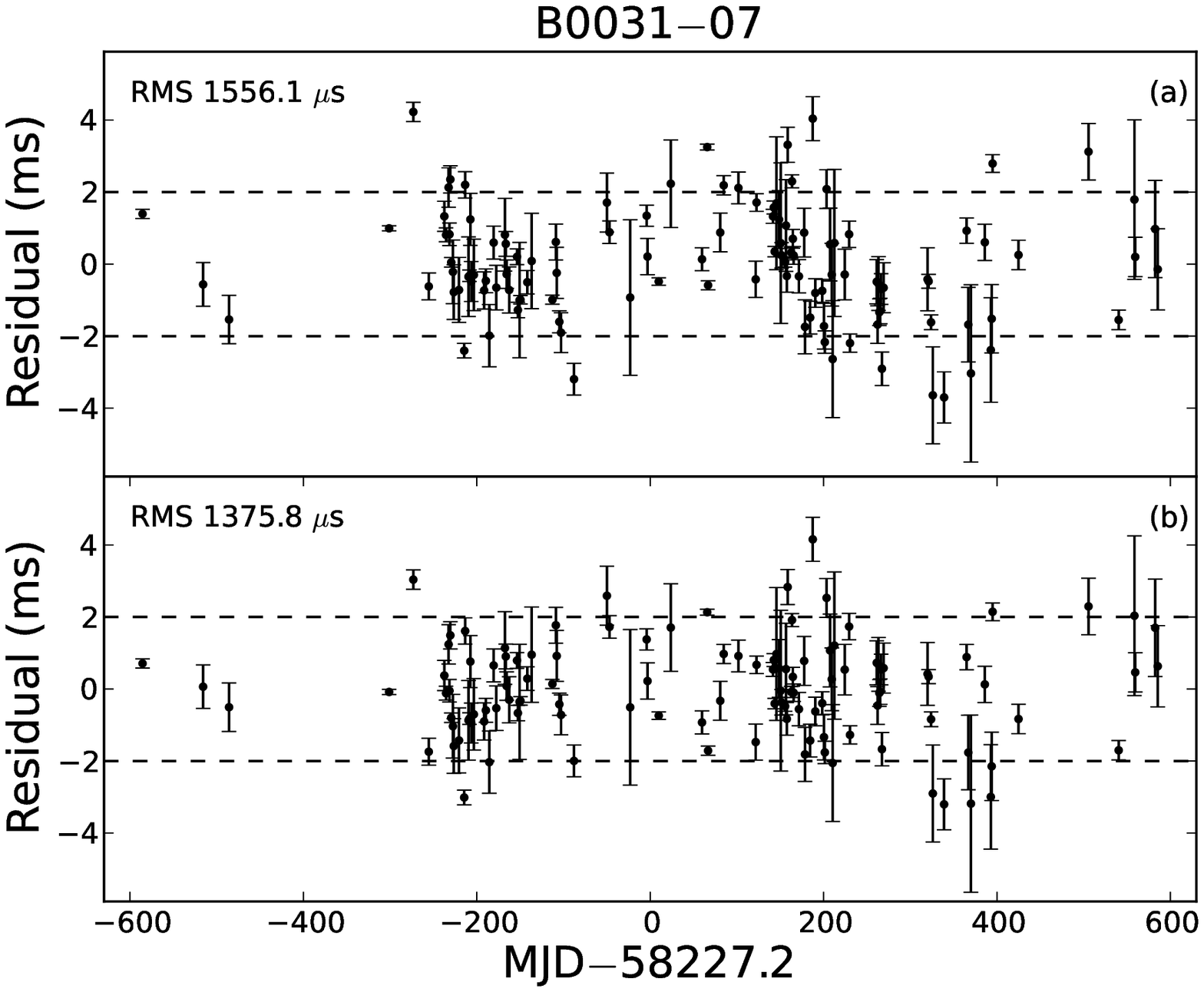}} &
\resizebox{0.45\hsize}{!}{\includegraphics[angle=0]{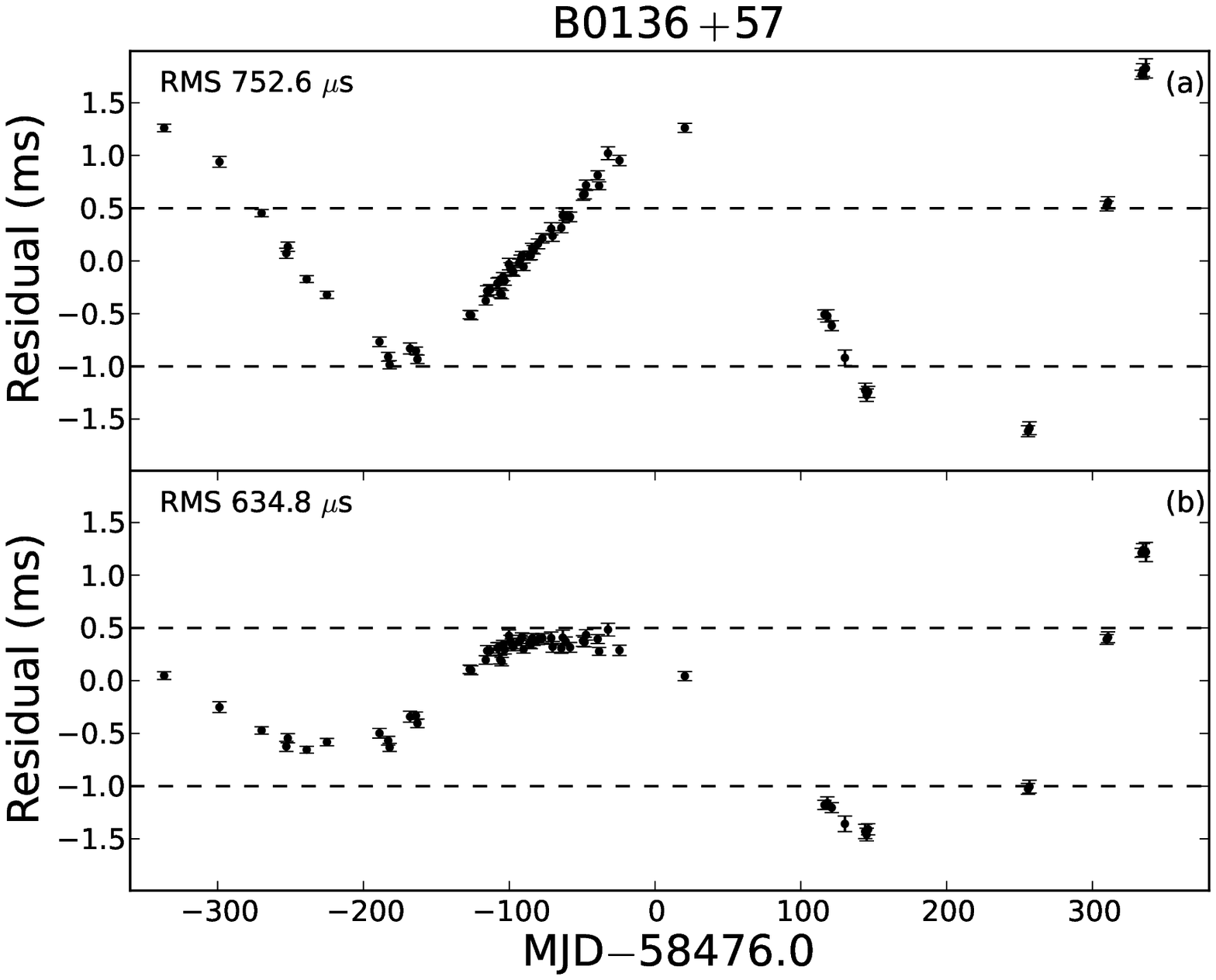}} \\
\resizebox{0.45\hsize}{!}{\includegraphics[angle=0]{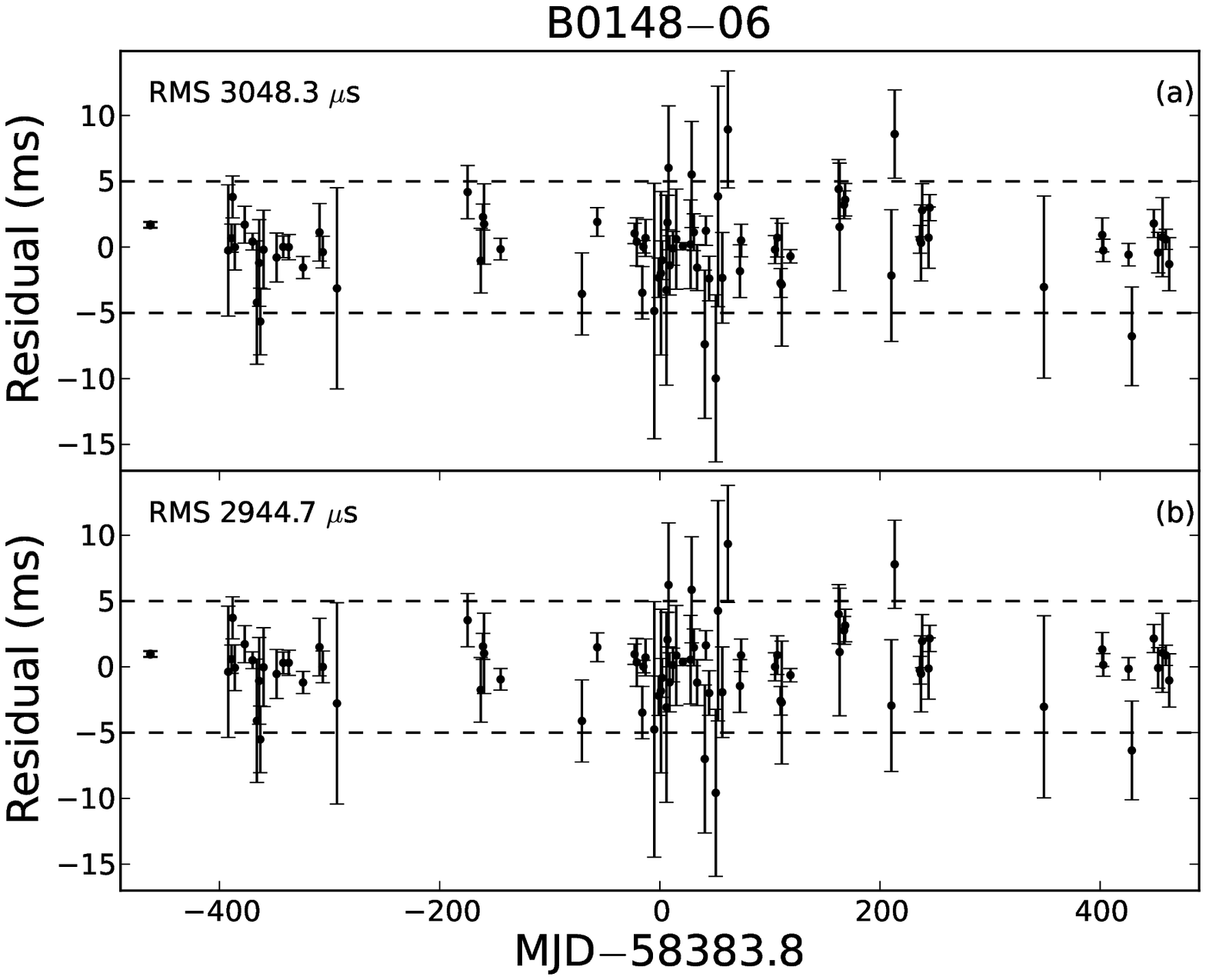}} &
\resizebox{0.45\hsize}{!}{\includegraphics[angle=0]{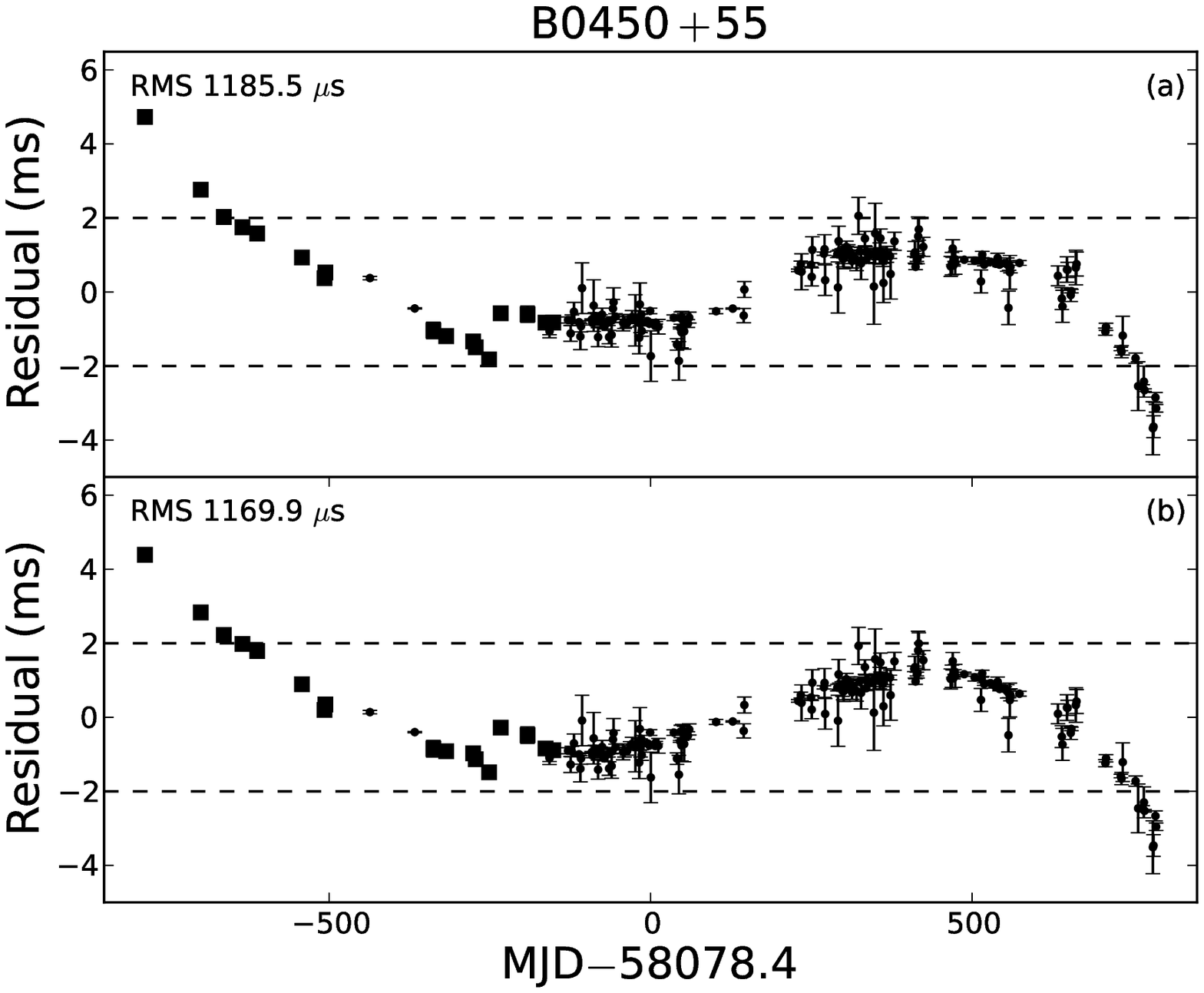}} \\
\resizebox{0.45\hsize}{!}{\includegraphics[angle=0]{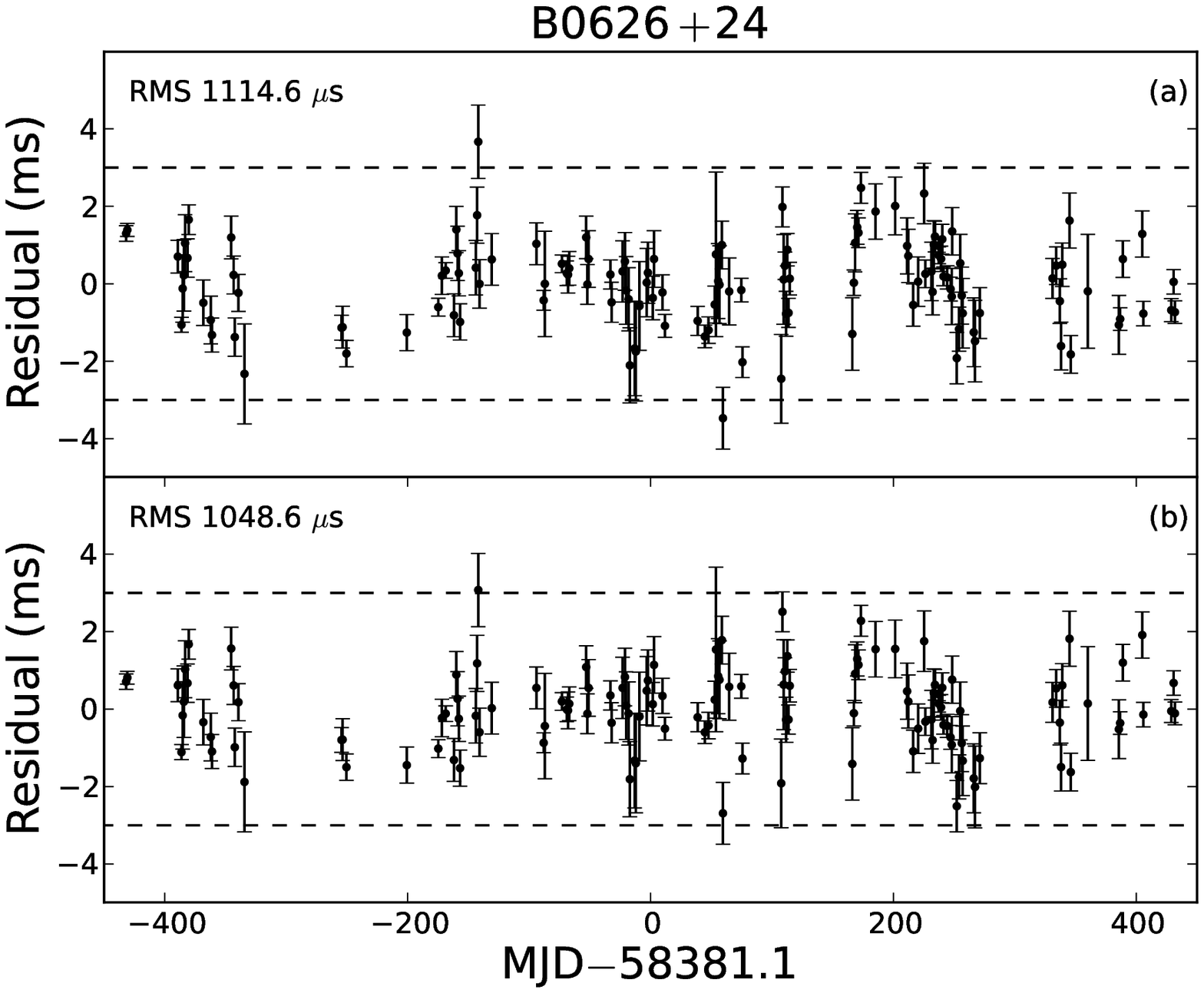}} &
\resizebox{0.45\hsize}{!}{\includegraphics[angle=0]{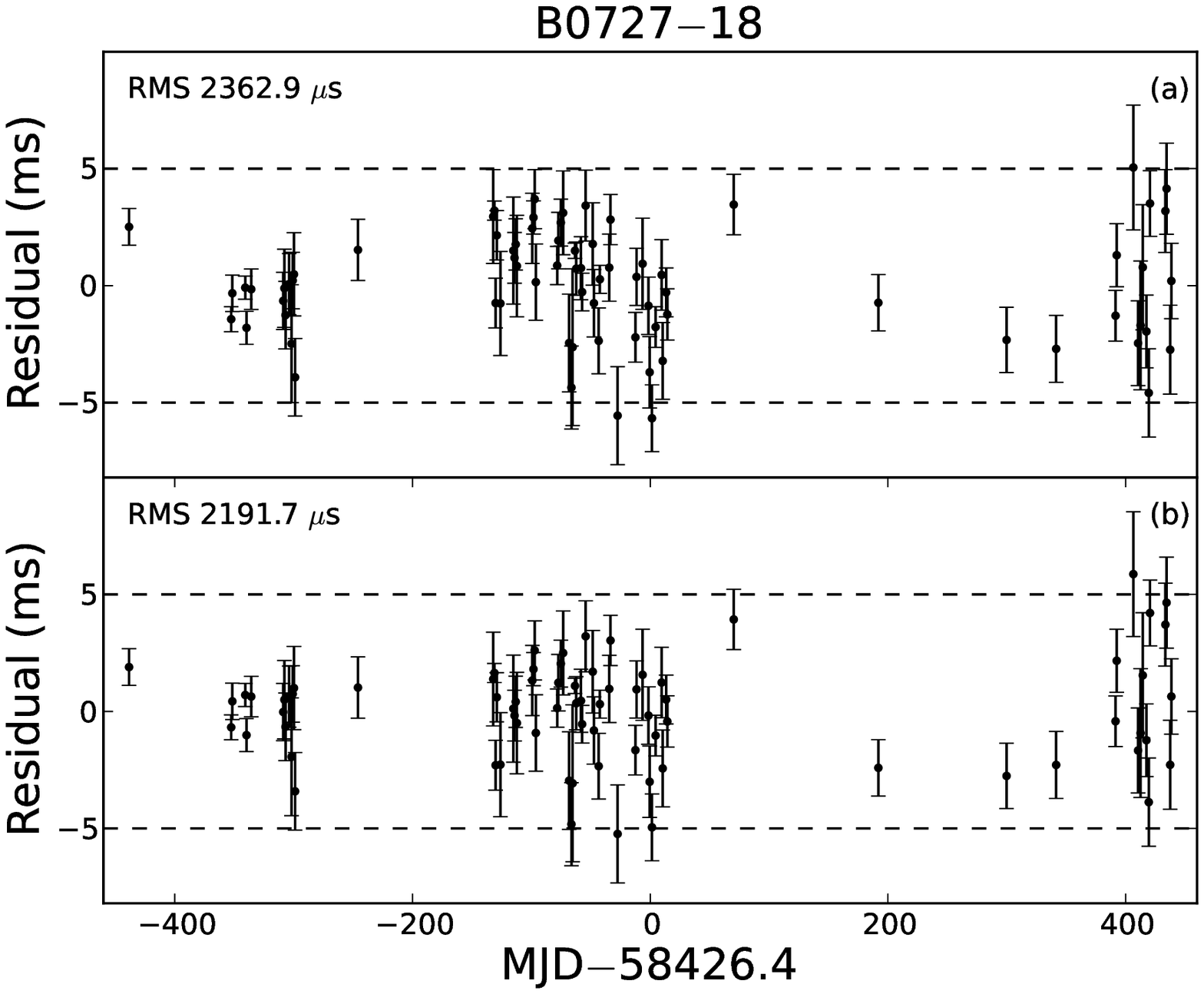}} \\
\resizebox{0.45\hsize}{!}{\includegraphics[angle=0]{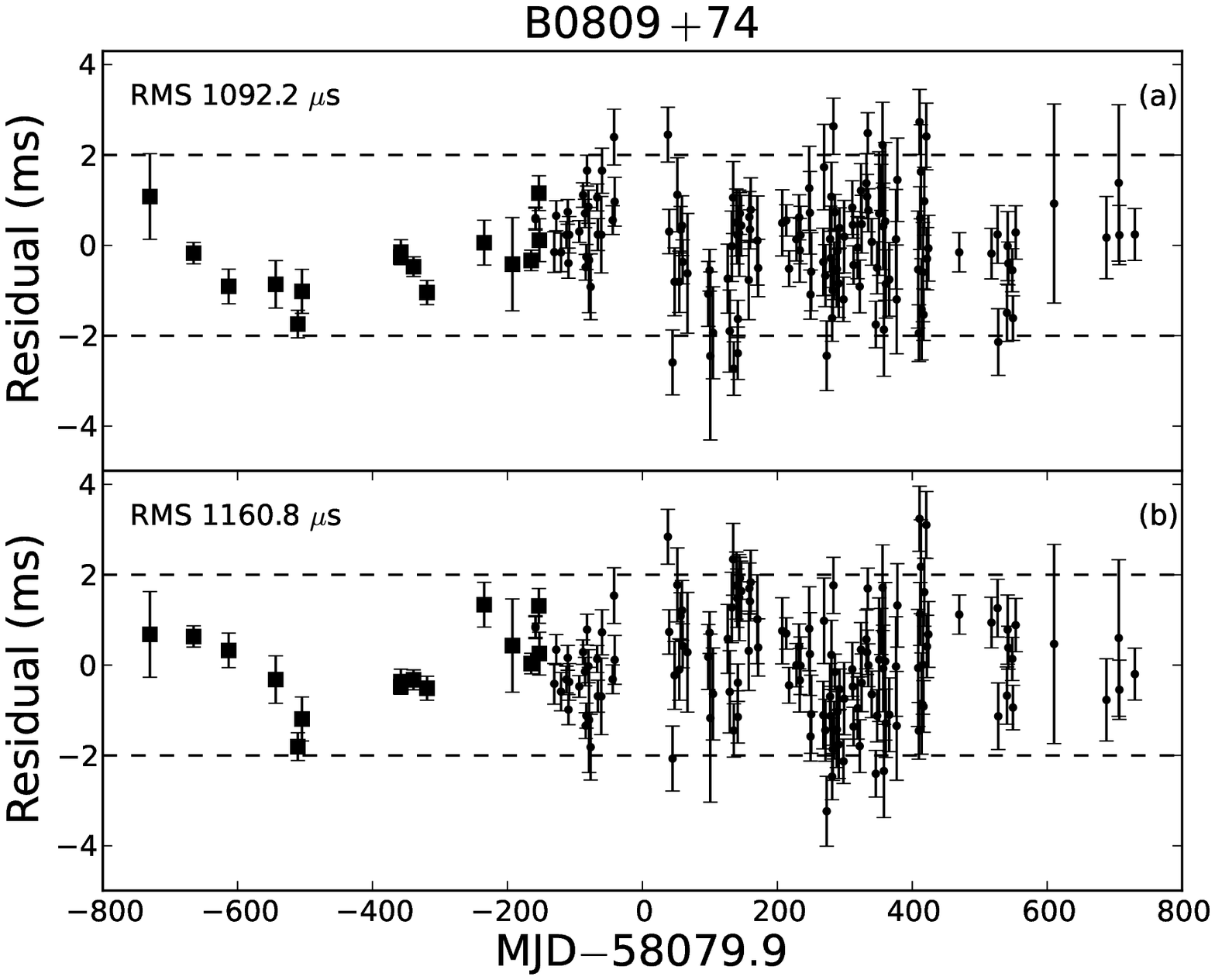}} &
\resizebox{0.45\hsize}{!}{\includegraphics[angle=0]{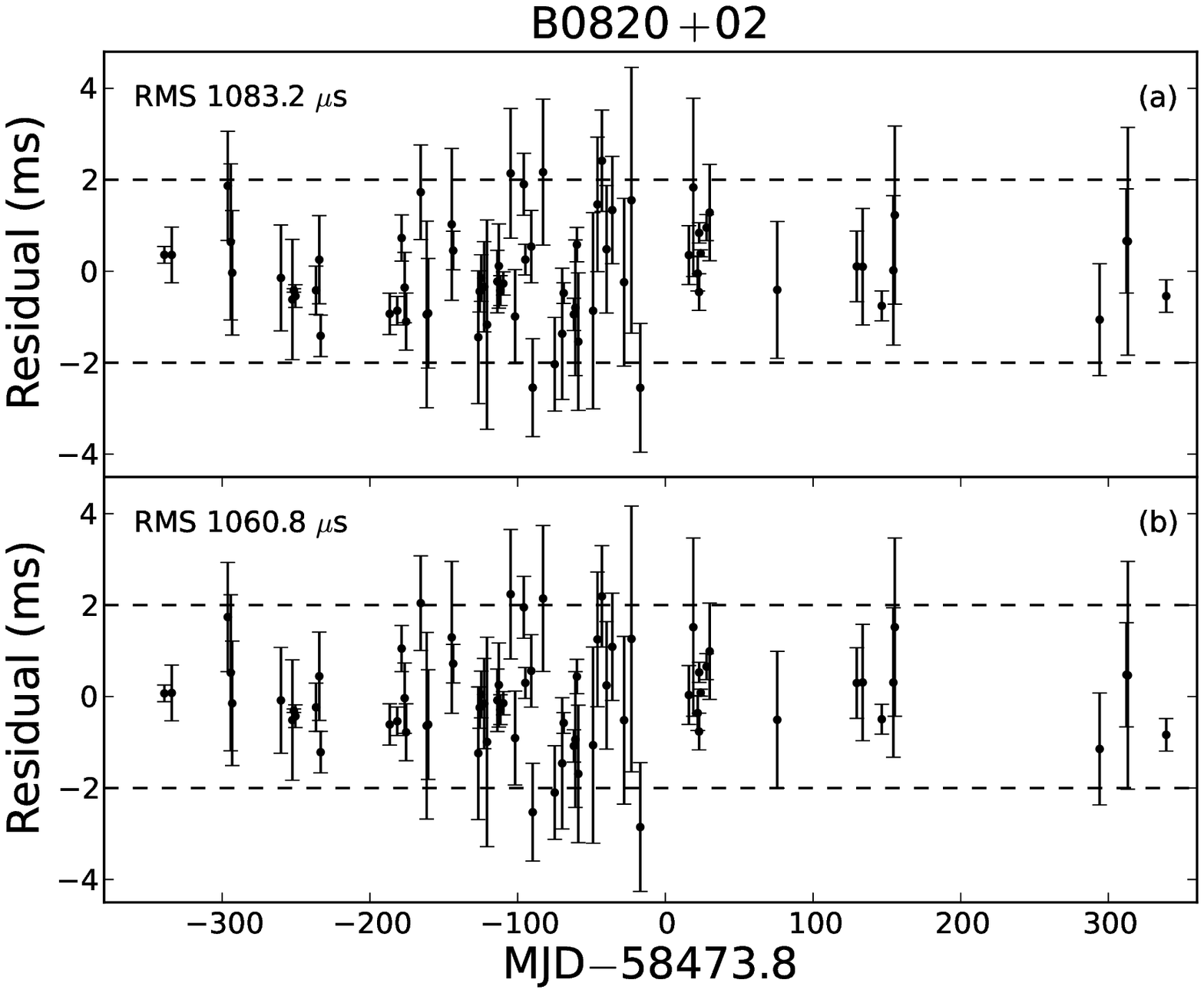}} \\
\end{tabular}
}
\end{center}
\caption{Timing residuals after fitting $\nu$ and $\dot{\nu}$ based on timing (a) and updated (b) ephemerides. Data points with dot and square symbols are at S and C$\textrm{-}$band, respectively. Axis scales of two panels are same for each pulsar.}
\label{fg:figure 4}
\end{figure*}

\addtocounter{figure}{-1}

\begin{figure*}[]
\addtocounter{subfigure}{2}
\begin{center}
\subfigure{
\begin{tabular}{cc}
\resizebox{0.45\hsize}{!}{\includegraphics[angle=0]{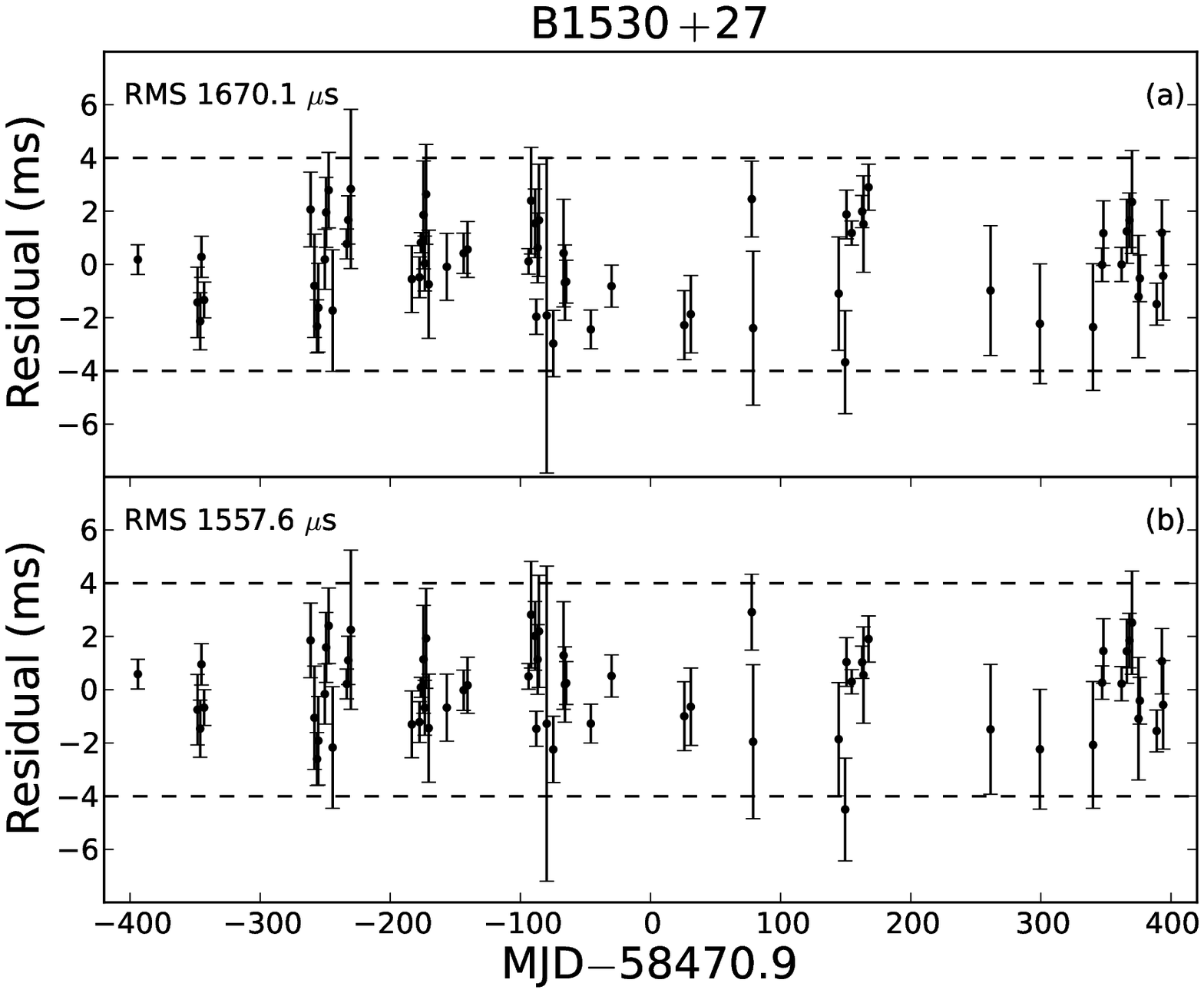}} &
\resizebox{0.45\hsize}{!}{\includegraphics[angle=0]{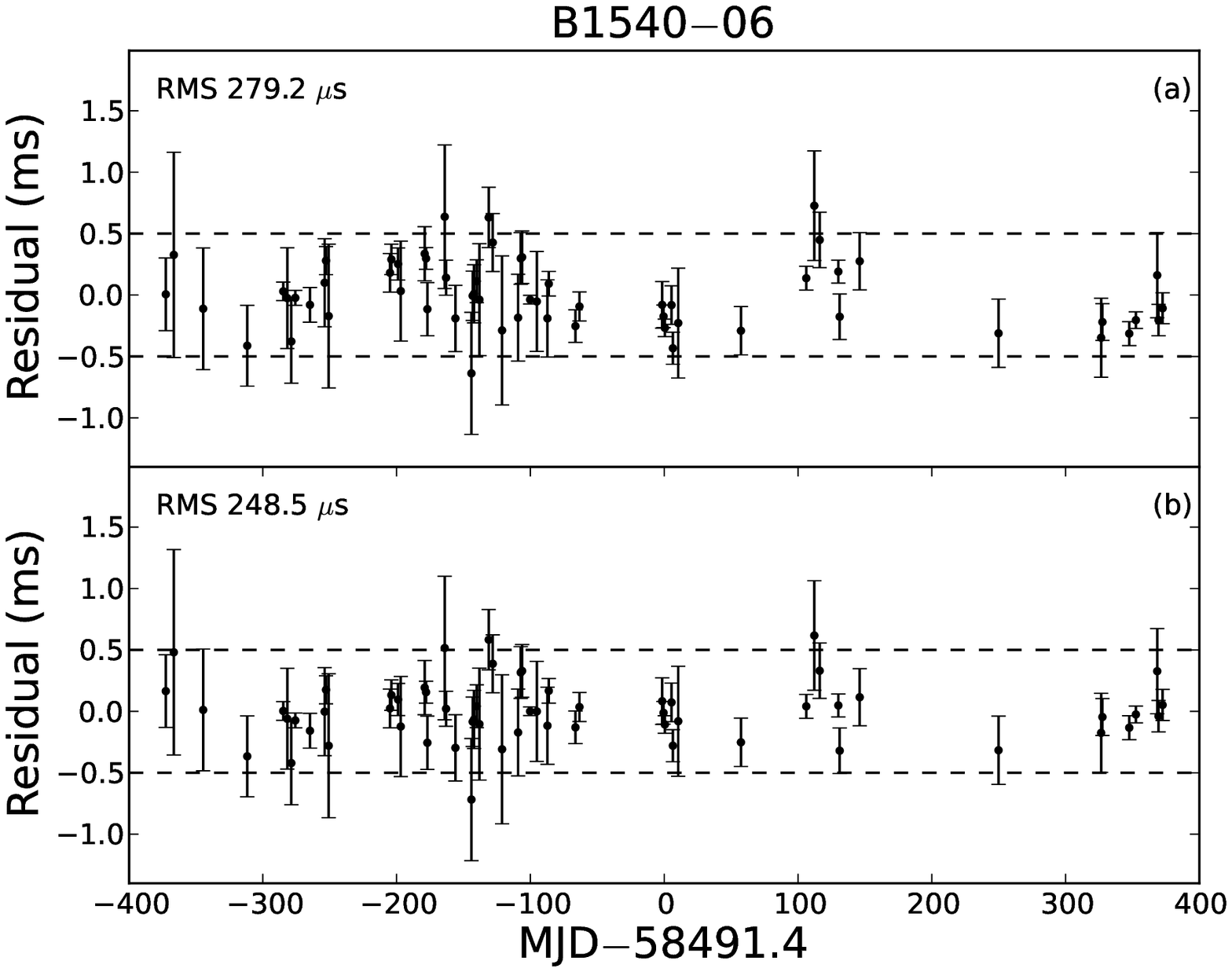}} \\
\resizebox{0.45\hsize}{!}{\includegraphics[angle=0]{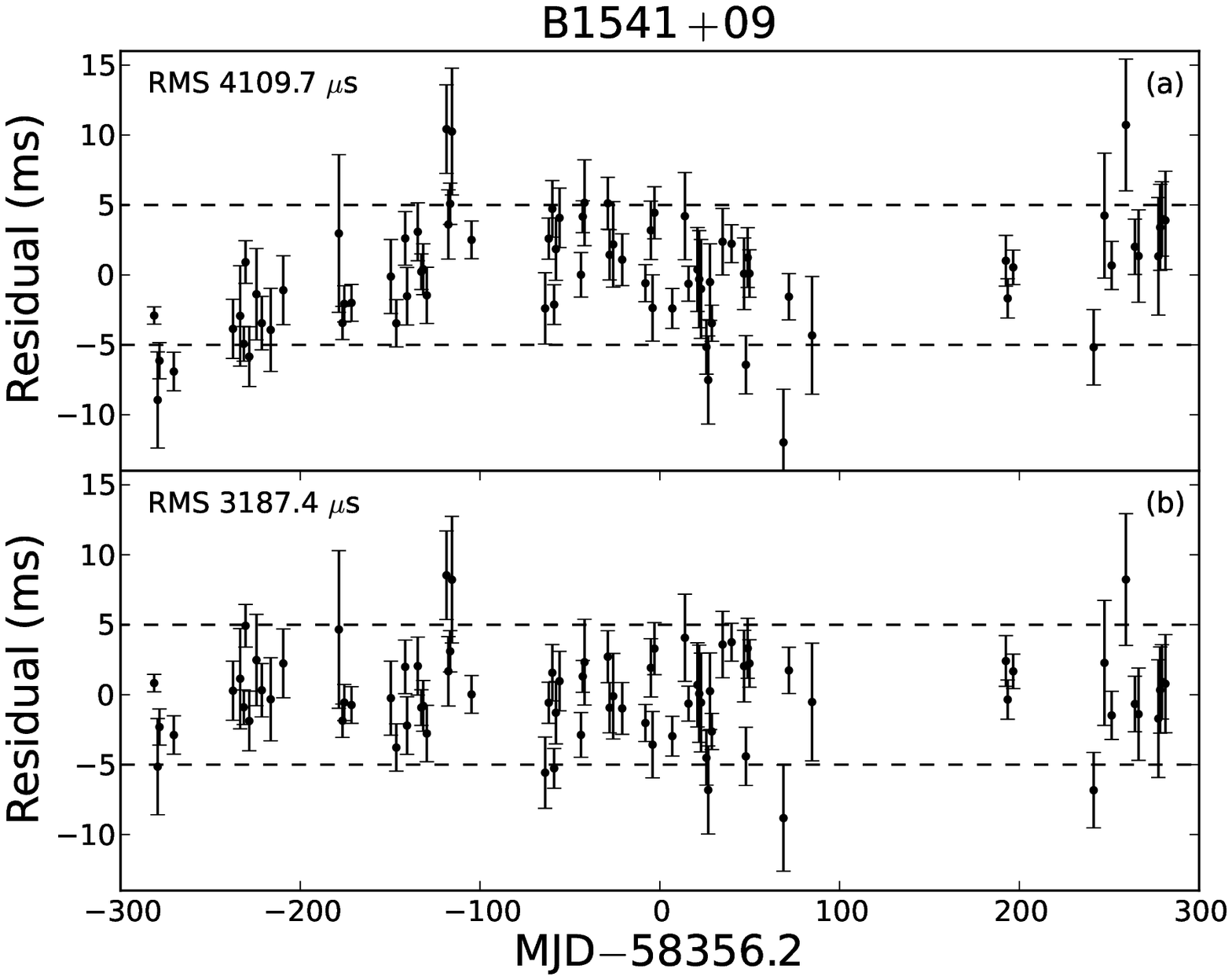}} &
\resizebox{0.45\hsize}{!}{\includegraphics[angle=0]{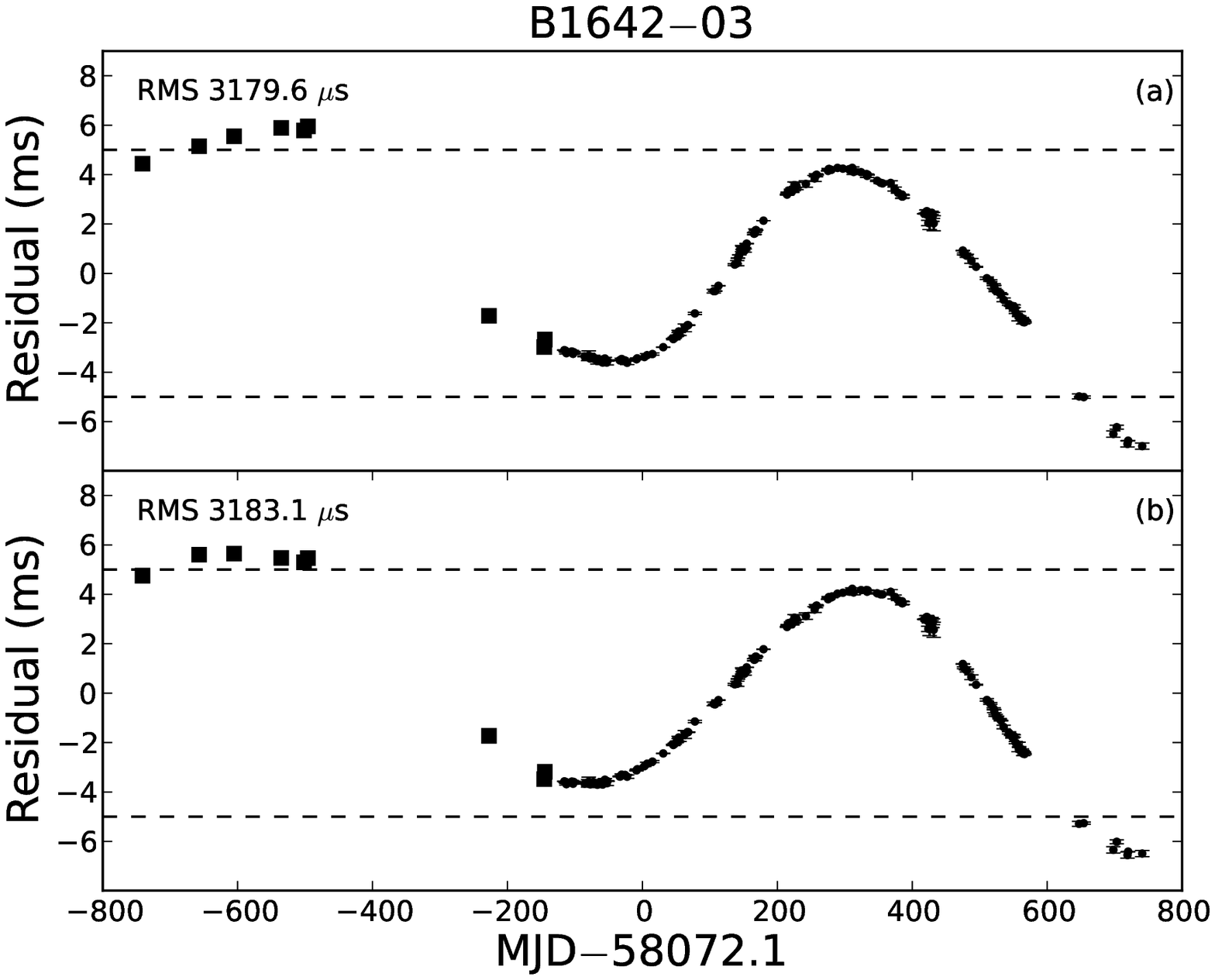}} \\
\resizebox{0.45\hsize}{!}{\includegraphics[angle=0]{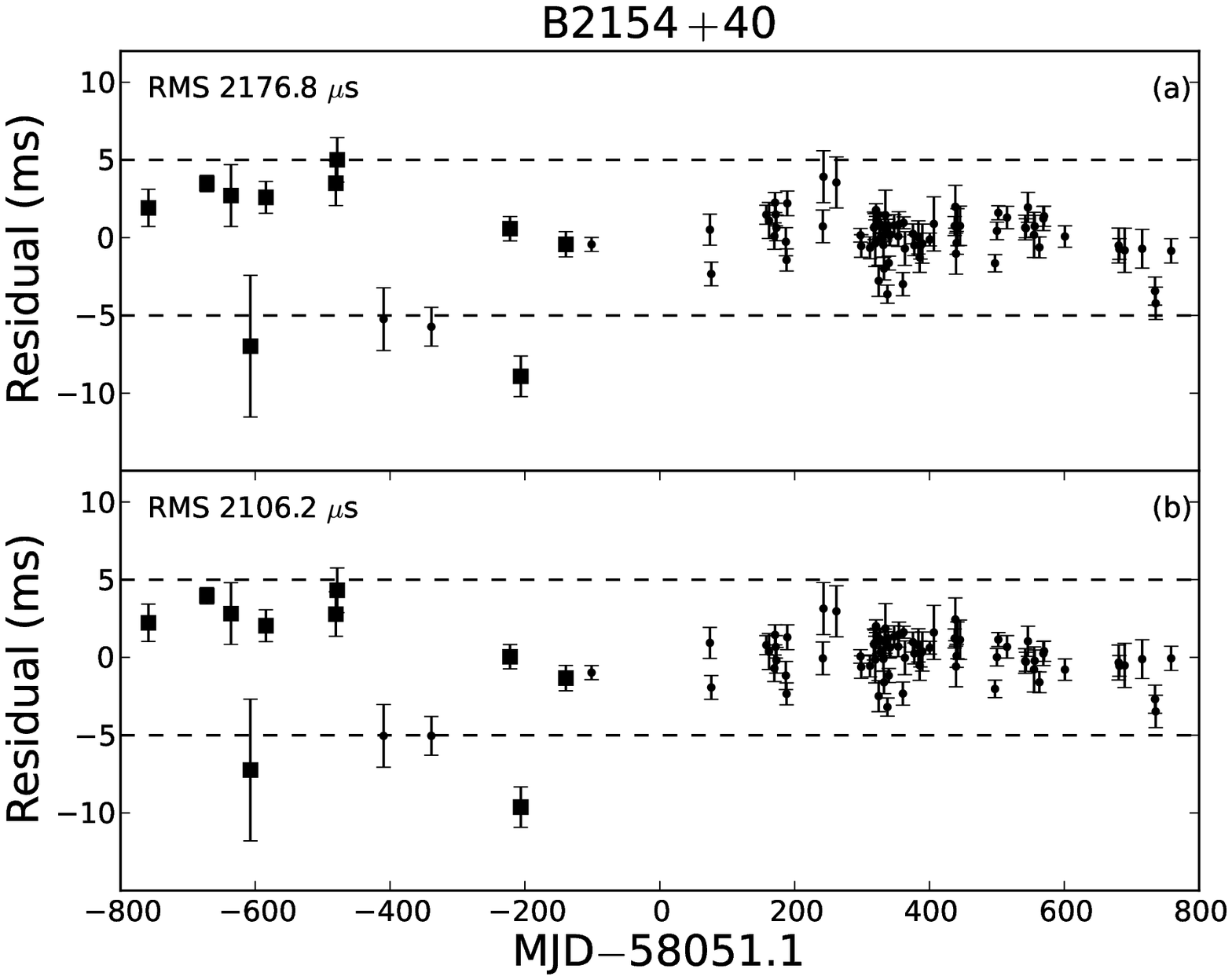}} &
\resizebox{0.45\hsize}{!}{\includegraphics[angle=0]{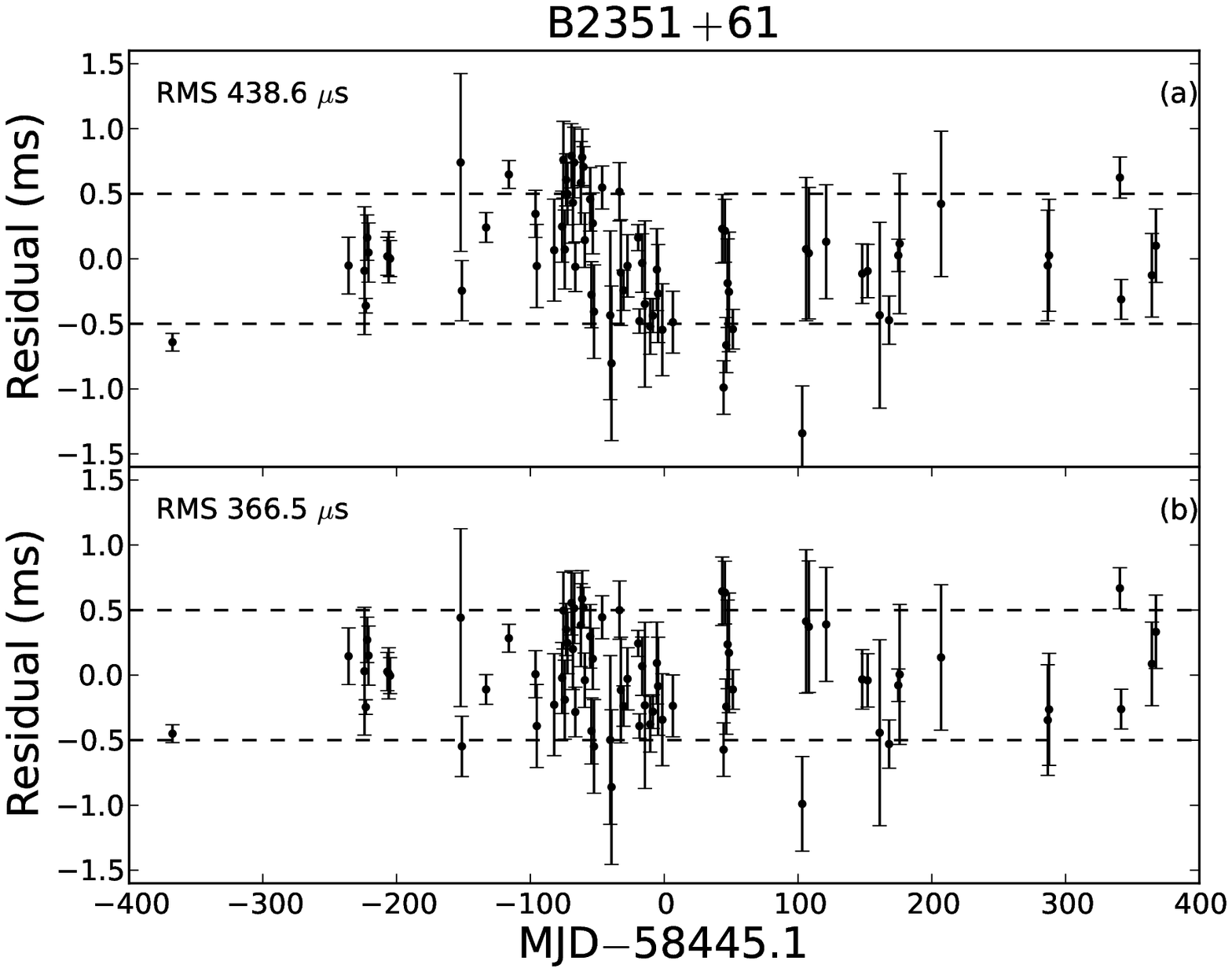}} \\
\end{tabular}
}
\end{center}
\caption{continued}
\label{fg:figure 4}
\end{figure*}

\begin{table*}[]
\footnotesize{
\caption{Inferred and fitted rotation parameters of 16 pulsars.}
\label{Tab:table 3}
\centering
\begin{tabular}{l l l l c c c c}
\hline
\hline
\multicolumn{1}{c}{Name} & Solutions & \multicolumn{1}{c}{$\nu$} & \multicolumn{1}{c}{$\dot{\nu}$} & \multicolumn{1}{c}{Epoch} & Epoch$_{\rm T}$ & Data span & No. of\\
 & &\multicolumn{1}{c}{(Hz)} & \multicolumn{1}{c}{(10$^{-15}$~s$^{-2}$)} &\multicolumn{1}{c}{(MJD)} & (MJD) & (MJD) & ToAs\\
\hline
\multirow{2}{*}{B0031$-$07} & Inferred & 1.06050003902(4)& $-$0.45889(7) & \multirow{2}{*}{58227} & \multirow{2}{*}{46635} & \multirow{2}{*}{57640$\textrm{-}$58814} & \multirow{2}{*}{117}\\
 & Fitted & 1.060500038930(7) & $-$0.4578(5) & & & & \\
\hline
\multirow{2}{*}{B0136$+$57} & Inferred & 3.670275200(9) & $-$144.29(2) & \multirow{2}{*}{58476} & \multirow{2}{*}{49289} & \multirow{2}{*}{58139$\textrm{-}$58813} & \multirow{2}{*}{66} \\
 & Fitted & 3.67027523674(2) & $-$144.152(3) & & & & \\
\hline
\multirow{2}{*}{B0148$-$06} & Inferred & 0.6827500310(1) & $-$0.2021(3) & \multirow{2}{*}{58383} & \multirow{2}{*}{49347} &  \multirow{2}{*}{57921$\textrm{-}$58848} & \multirow{2}{*}{81} \\
 & Fitted  & 0.68275002918(1) & $-$0.2045(8) & & & & \\
\hline
\multirow{2}{*}{B0154$+$61} & Inferred & 0.4251905242(6) &  $-$34.176(2) & \multirow{2}{*}{58400} & \multirow{2}{*}{49709} &\multirow{2}{*}{58058$\textrm{-}$58742} & \multirow{2}{*}{136}\\
 & Fitted & 0.425190536230(5) & $-$34.1756(6) & & & & \\
\hline
\multirow{2}{*}{B0450$+$55} & Inferred & 2.934865512(4) & $-$20.56(1) & \multirow{2}{*}{58078} & \multirow{2}{*}{49910} &\multirow{2}{*}{57292$\textrm{-}$58866} & \multirow{2}{*}{188}\\
 & Fitted & 2.93486550969(1) & $-$20.6703(5) & & & & \\
\hline
\multirow{2}{*}{B0611$+$22} & Inferred & 2.98503223(7) & $-$524.8(2) & \multirow{2}{*}{58435} & \multirow{2}{*}{49674} & \multirow{2}{*}{58130$\textrm{-}$58741} & \multirow{2}{*}{102}\\
 & Fitted & 2.985032106(5)  & $-$526.2(4) & & & & \\
\hline
\multirow{2}{*}{B0626$+$24} & Inferred & 2.0980882323(8) & $-$8.791(2) & \multirow{2}{*}{58381} & \multirow{2}{*}{49438} & \multirow{2}{*}{57949$\textrm{-}$58813} & \multirow{2}{*}{131}\\
 & Fitted & 2.098088239948(10) & $-$8.7795(9) & & & & \\
\hline
\multirow{2}{*}{B0727$-$18} & Inferred & 1.960113351(5) & $-$72.56(1) & \multirow{2}{*}{58426} & \multirow{2}{*}{49720} &\multirow{2}{*}{57987$\textrm{-}$58866} & \multirow{2}{*}{75}\\
 & Fitted & 1.96011325712(3) & $-$72.868(2) & & & & \\
\hline
\multirow{2}{*}{B0809$+$74} & Inferred & 0.77384911485(4) & $-$0.1006(1) & \multirow{2}{*}{58079}  & \multirow{2}{*}{49162} &\multirow{2}{*}{57349$\textrm{-}$58810} & \multirow{2}{*}{153}\\
 & Fitted & 0.773849114816(3) & $-$0.1005(2) & & & & \\
\hline
\multirow{2}{*}{B0820$+$02} & Inferred & 1.15623927397(5) & $-$0.1404(1) & \multirow{2}{*}{58473}  & \multirow{2}{*}{49281} &\multirow{2}{*}{58134$\textrm{-}$58813} & \multirow{2}{*}{71}\\
 & Fitted & 1.15623927402(1) & $-$0.136(1) & & & & \\
\hline
\multirow{2}{*}{B1530$+$27} & Inferred & 0.8890182224(3) & $-$0.6168(6) & \multirow{2}{*}{58470}  & \multirow{2}{*}{49666} &\multirow{2}{*}{58076$\textrm{-}$58866} & \multirow{2}{*}{64}\\
 & Fitted & 0.889018222765(9) & $-$0.6147(10) & & & & \\
\hline
\multirow{2}{*}{B1540$-$06} & Inferred & 1.4103084243(4) & $-$1.7391(8) & \multirow{2}{*}{58491}  & \multirow{2}{*}{49423} &\multirow{2}{*}{58118$\textrm{-}$58864} & \multirow{2}{*}{60}\\
 & Fitted & 1.410308423169(3) & $-$1.7489(3) & & & & \\
\hline
\multirow{2}{*}{B1541$+$09} & Inferred & 1.3360967749(3) & $-$0.7836(7) & \multirow{2}{*}{58356}  & \multirow{2}{*}{48716} &\multirow{2}{*}{58075$\textrm{-}$58638} & \multirow{2}{*}{79}\\
 & Fitted & 1.33609677905(4) & $-$0.784(5) & & & & \\
\hline
\multirow{2}{*}{B1642$-$03} & Inferred & 2.5793706170(7) & $-$11.839(1) & \multirow{2}{*}{58072}  & \multirow{2}{*}{46515} &\multirow{2}{*}{57331$\textrm{-}$58813} & \multirow{2}{*}{156}\\
 & Fitted & 2.57937061296(3) & $-$11.837(1) & & & & \\
\hline
\multirow{2}{*}{B2154$+$40} & Inferred & 0.6556223737(2) & $-$1.5075(5) & \multirow{2}{*}{58051}  & \multirow{2}{*}{49277} &\multirow{2}{*}{57292$\textrm{-}$58810} & \multirow{2}{*}{88}\\
 & Fitted & 0.655622382900(5) & $-$1.4762(3) & & & & \\
\hline
\multirow{2}{*}{B2351$+$61} & Inferred & 1.0584288921(6) & $-$18.200(1) & \multirow{2}{*}{58445}  & \multirow{2}{*}{49405} &\multirow{2}{*}{58077$\textrm{-}$58813} & \multirow{2}{*}{74}\\
 & Fitted & 1.058428884927(4) & $-$18.2220(4) & & & & \\
\hline
\end{tabular}
\\
Note: Epoch$_{\rm T}$ is the epoch of timing rotation solution in \cite{hlk+04}.
}
\end{table*}

For the convenience in comparing previous rotation solutions in \cite{hlk+04} with corresponding solutions derived in our work, we also inferred the $\nu$ and $\dot{\nu}$ in \cite{hlk+04} to the epoch of our rotation solutions.  Table~\ref{Tab:table 3} presents fitted and inferred rotation parameters, epoch of solutions, data span and number of ToAs of 16 pulsars.

\section{Discussion}
\label{sect:discussion}

With VLBI astrometric parameter solutions, the effects caused by inaccurate astrometric parameters can be properly removed in pulsar timing residuals. Residual variations differ a lot for different pulsars after updating astrometric parameters using the latest VLBI solutions. For PSR B0611$+$22 that has loud timing noise, there is a large difference of its declination between timing and VLBI solutions. However, its differences between timing residuals based on two kinds of ephemerides are imperceptible compared with timing noise (see Figure~\ref{fg:figure 3}). For some pulsars without significant timing irregularities, there are obvious differences between residuals based on timing and updated ephemerides (see Figure~\ref{fg:figure 4}). PSR B1541$+$09 has big differences of declination and $\mu_{\delta}$ between timing and VLBI solutions as shown in Table~\ref{Tab:table 1}. Its timing residuals based on timing ephemeris are apparently different from residuals based on updated ephemeris. Differences between timing residuals based on two kinds of ephemerides are less apparent for pulsars that have small differences of astrometric parameters between timing and VLBI solutions, like PSR B2351$+$61. Compared with PSR B1541$+$09, PSR B0626$+$24 has larger differences of declination and $\mu_{\delta}$ between two solutions. However, its timing residuals based on two kinds of ephemerides have less obvious differences. This is probably because of big ToA errors of PSR B0626$+$24. ToA errors of PSRs B0136$+$57 and B1642$-$03 are much smaller than that of PSR B0626$+$24. For these two pulsars, declination and proper motion differences between two solutions are much smaller than that of PSR B0626$+$24, but differences between timing residuals based on two kinds of ephemerides are more obvious than residual differences of PSR B0626$+$24.

Timing residual fluctuations caused by inaccurate astrometric parameters influence measurements of timing irregularities. Benefiting from the independence and high accuracy of astrometric parameters measured by the VLBI, it is possible to perform more intrinsic measurements of timing irregularities. PSR B0154$+$61 suffered one glitch around MJD 58266.4 as shown in panel (a) of Figure~\ref{fg:figure 1}. The proper motion solution of PSR B0154$+$61 obtained with timing observations has an obvious difference with the VLBI solution. Errors of timing proper motion are much larger than that of VLBI proper motion solution. Due to inaccurate astrometric parameters, residuals after fitting glitch parameters based on timing ephemeris still have obvious fluctuations as shown in panel (b). Glitch parameters measured with timing ephemeris are different from that measured with updated ephemeris. As shown in Table~\ref{Tab:table 2}, glitch epoch measured with timing ephemeris is about 13~d later than that measured with updated ephemeris. Differences of $\Delta \nu$ and $\Delta \dot{\nu}$ between two solutions are 3$\times 10^{-10}$~Hz and $-$1.5$\times 10^{-17}$~s$^{-2}$, respectively. In addition, errors of parameters measured with timing ephemeris are larger than that measured with updated ephemeris.

Pulsar rotation behaviours can be described well by simple spin$\textrm{-}$down models. However, apparent biases may appear over long time, especially when timing irregularities exist. So that, it is necessary to refit rotation parameters using the latest timing data. Correct rotation solutions are derived on the premise of accurate astrometric parameters. Timing residuals based on timing ephemerides may have obvious fluctuations caused by inaccurate astrometric parameters, like residuals of PSR B1541$+$09 shown in panel (a) of corresponding plot in Figure~\ref{fg:figure 4}. While, none of these fluctuations are remained in timing residuals based on updated ephemerides. Hence, VLBI astrometric parameter solutions have important meaning for obtaining accurate rotation parameters. Comparing fitted $\nu$ and inferred $\nu$ in Table~\ref{Tab:table 3}, they have obvious differences, especially for pulsars with large $\left|\dot{\nu} \right|$. Most of differences between the fitted and the inferred values of $\nu$ range from 10$^{-9}$ to 10$^{-8}$~Hz. Errors of fitted $\nu$ are much smaller than that of inferred $\nu$ for all pulsars. In most cases, differences between fitted and inferred $\dot{\nu}$ are larger than 10$^{-18}$~s$^{-2}$. For pulsars with smaller values of $\left|\dot{\nu} \right|$, differences of $\dot{\nu}$ between two solutions are smaller than other pulsars. What's more, errors of fitted $\dot{\nu}$ are larger than that of inferred $\dot{\nu}$ for these pulsars. This is probably because time spans of our timing data are relatively not long enough to precisely measure $\dot{\nu}$.

\begin{landscape}
\begin{table}[tb]
\footnotesize{
\caption{Parameters relative to the Shklovsky effect and newly calculated characteristic ages of 16 pulsars.}
\label{Tab:table 4}
\centering
\begin{tabular}{l l l l l l l l l l}
\hline
\hline
\multicolumn{1}{c}{Name} &
\multicolumn{1}{c}{$V_{\rm T}$} &
\multicolumn{1}{c}{$d$} &
\multicolumn{1}{c}{$P$} &
\multicolumn{1}{c}{$\dot{P}_{\rm m}$} &
\multicolumn{1}{c}{$\dot{P}_{\rm s}$} &
\multicolumn{1}{c}{$\dot{P}_{\rm s}/\dot{P}_{\rm m}$} &
\multicolumn{1}{c}{$\tau_{\rm m}$} &
\multicolumn{1}{c}{$\tau_{\rm i}$} &
\multicolumn{1}{c}{$\tau_{\rm p}$} \\
 &
\multicolumn{1}{c}{(km s$^{-1}$)} &
\multicolumn{1}{c}{(kpc)} &
\multicolumn{1}{c}{(s)} &
\multicolumn{1}{c}{(10$^{-15}$~s~s$^{-1}$)} &
\multicolumn{1}{c}{(10$^{-19}$~s~s$^{-1}$)}  &
\multicolumn{1}{c}{(\%)}  &
\multicolumn{1}{c}{(Myr)} &
\multicolumn{1}{c}{(Myr)} &
\multicolumn{1}{c}{(Myr)}
\\
\hline
B0031$-$07 & 77.00$^{\rm b}$    & 1.26$^{\rm b}$ & 0.942951403386(6) & 0.4070(4)  & 4.79323268066(3) & 0.12 & 36.73(4)   & 36.78(4)   & 36.6243(6)\\
B0136$+$57 & 319.0$^{\rm b}$    & 2.65$^{\rm b}$ & 0.272459130582(2) & 10.7010(2) & 11.30225636215(7) & 0.01 & 0.40368(1) & 0.40373(1) & 0.403268(5)\\
B0148$-$06 & 261.0$^{\rm c}$    & 4.60$^{\rm c}$ & 1.46466489529(2)  & 0.437(2)   & 23.4309420922(4) & 0.53 & 53.1(2)    & 53.4(2)   & 52.469(6)\\
B0154$+$61 & 383.7$^{\rm c}$    & 1.80$^{\rm c}$ & 2.35188677732(3) & 189.038(3) & 207.804907222(2) & 0.01 & 0.197256(3) & 0.197278(3) & 0.1973619(5)\\
B0450$+$55 & 317.0$^{\rm b}$    & 1.19$^{\rm b}$ & 0.340731115855(1) & 2.39977(5) & 31.0821925829(1) & 0.13 & 2.25115(5) & 2.25407(5)  & 2.27724(7)\\
B0611$+$22 & 21.00$^{\rm c}$    & 3.55$^{\rm c}$ & 0.33500477221(4) & 59.05(4)   & 0.044956192797(5) & 0    & 0.08995(7) & 0.08995(7) & 0.089332(2)\\
B0626$+$24 & 84.00$^{\rm c}$    & 3.00$^{\rm c}$ & 0.476624376878(2) & 1.9944(2)  & 1.210993642508(6) & 0.01 & 3.7890(4)  & 3.7892(4)  & 3.78649(6) \\
B0727$-$18 & 179.9$^{\rm c}$    & 2.04$^{\rm c}$ & 0.510174601578(7) & 18.9659(6) & 8.7433773083(1) & 0    & 0.42649(1) & 0.42651(1) & 0.426675(4)\\
B0809$+$74 & 102.0$^{\rm a}$    & 0.43$^{\rm a}$ & 1.292241576368(5) & 0.1679(3)  & 33.7756638012(1) & 2.01 & 122.1(2)   & 124.6(2)  & 121.87(1)\\
B0820$+$02 & 47.80$^{\rm c}$    & 2.66$^{\rm c}$ & 0.864872887877(9) & 0.1014(9)  & 0.802517912658(8) & 0.08 & 135.2(12)  & 135.3(12) & 131.157(8)\\
B1530$+$27 & 144.0$^{\rm c}$    & 1.60$^{\rm c}$ & 1.12483633563(1) & 0.778(1)   & 15.7478994453(2) & 0.2  & 22.93(4)   & 22.98(4) &  22.877(2)\\
B1540$-$06 & 247.4$^{\rm c}$    & 3.11$^{\rm c}$ & 0.709064757447(1) & 0.8793(1)  & 15.07485676089(3) & 0.17 & 12.785(2)  & 12.807(2) & 12.7818(6)\\
B1541$+$09 & 277.0$^{\rm b}$    & 7.20$^{\rm b}$ & 0.74844877682(2) & 0.439(3)   & 8.6162327085(2) & 0.2  & 27.02(17)  & 27.07(17) & 27.438(3)\\
B1642$-$03 & 386.4$^{\rm c}$    & 3.97$^{\rm c}$ & 0.387691475966(5) & 1.7791(2)  & 15.7506425803(2) & 0.09 & 3.4550(4)  & 3.4581(4) & 3.45241(3)\\
B2154$+$40 & 264.0$^{\rm b}$    & 3.40$^{\rm b}$ & 1.52526824294(1) & 3.4342(7)  & 33.7756301241(3) & 0.1  & 7.042(1)   & 7.049(1) & 7.0452(1)\\
B2351$+$61 & 268.0$^{\rm c}$    & 2.42$^{\rm c}$ & 0.944796588832(3) & 16.2657(3) & 30.2915037893(1) & 0.02 & 0.92094(2) & 0.92111(2) & 0.921087(5)\\
\hline
\end{tabular}
\\
Reference: a. \cite{bbgt02}; b. \cite{cbv+09}; c. \cite{dgb+19}.
Note: $\tau_{\rm i}$ is the intrinsic age with the Shklovsky effect subtracted. $\tau_{\rm p}$ is calculated based on previous rotation solutions in \cite{hlk+04}.
}
\end{table}
\end{landscape}

According to \cite{shk70}, a large transverse velocity of pulsar will cause an apparent Doppler shift in rotational frequency which is like the ``train$\textrm{-}$whistle" effect, giving rise to the increase in period derivative $\dot{P}$
\begin{equation}
  \dot{P}_{\rm s}/P = \frac{V_{\rm T}^{2}}{cd},
\label{eq:equation 3}
\end{equation}
where $V_{\rm T}$ and $c$ are transverse velocity to the line of sight and speed of light in vacuum, respectively. Given accurate pulsar distance $d$ (kpc) and proper motion $\mu$ (mas~yr$^{-1}$), $V_{\rm T}$ (km~s$^{-1}$) can be obtained by $V_{\rm T} = 4.74\mu d$ \citep{ll94}. Due to the existence of the Shklovsky effect, the relation between measured $\dot{P}_{\rm m}$ and intrinsic spin$\textrm{-}$down rate $\dot{P}_{\rm i}$ is \citep{cam94}
\begin{equation}
  \frac{\dot{P}_{\rm m}}{P} \approx \frac{\dot{P}_{\rm i}}{P} + \frac{V_{\rm T}^{2}}{cd} \equiv \frac{\dot{P}_{\rm i}}{P} + \frac{\dot{P}_{\rm s}}{P},
\label{eq:equation 4}
\end{equation}
We calculated $\dot{P}_{\rm s}$, characteristic age $\tau_{\rm m}$ and intrinsic age $\tau_{\rm i}$ of these 16 pulsars. The results are listed in Table~\ref{Tab:table 4}. Comparing $\dot{P}_{\rm s}$ with $\dot{P}_{\rm m}$ in Table~\ref{Tab:table 4}, the Shklovsky effect has no significant influence on the $\dot{P}$ determination for most normal pulsars except some nearby pulsars. PSR B0809$+$74 is a nearby pulsar with the distance of 0.43~kpc. The rotational period and transverse velocity of this pulsar are 1.292~s and 102~km~s$^{-1}$, respectively. For this pulsar, the corresponding $\dot{P}_{\rm s}$ is about 3.378$\times 10^{-18}$~s~s$^{-1}$, which amounts to about 2\% of the $\dot{P}_{\rm m}$. Pulsar characteristic ages increase after subtracting the Shklovsky effect. For most pulsars, increments are not obvious except PSR B0809$+$74 whose age increment is 2.5~Myr. The $\tau_{\rm p}$ of 16 pulsars in the last column are calculated based on previous rotation solutions in \cite{hlk+04}. For pulsars like PSR B0809$+$74 and PSR B1540$-$06, increments of characteristic age after subtracting the Shklovsky effect are much larger than differences between the $\tau_{\rm m}$ and $\tau_{\rm p}$. So that, the Shklovsky effect is significant in these cases, even if age increments are not obvious compared with the $\tau_{\rm m}$. Influence of the Shklovsky effect on $\dot{P}$ and $\tau_{\rm c}$ can be better estimated with VLBI astrometric parameter solutions.

VLBI astrometric parameter solutions are independent and highly accurate. They can bring obvious optimizations for pulsar timing as discussed above. Hence, it has important meaning to measure pulsar astrometric parameters with the VLBI, especially for pulsars with timing irregularities. Besides the Very Long Baseline Array (VLBA) \citep{nbc+94}, the European VLBI Network (EVN) \citep{boo91} and the Australian Long Baseline Array (LBA) \citep{nor88}, the East Asia VLBI Network (EAVN) \citep{wha+16} has great potential to perform pulsar observations. The longest baselines of the VLBA, European part of the EVN and the LBA are 8600~km\footnote{https://science.nrao.edu/facilities/vlba/introduction-to-the-VLBA}, 7139~km\footnote{https://www.evlbi.org/capabilities} and 1400~km, respectively. For the EAVN, the maximum of baseline is 5000~km. The EAVN has the ability to conduct observations at frequencies as low as 1.6~GHz. It consists as many as 21 telescopes. Part of these telescopes are in the Japanese VLBI Network (JVN) \citep{ok96}. Previously, the JVN had successfully performed experiments on giant radio pulses of the Crab pulsar \citep{ttk+16}. Besides, many telescopes in China have ability to carry out pulsar observations. So that, it's expectable to obtain accurate pulsar astrometric solutions by the EAVN.

\section{Conclusion}
\label{sect:conclusion}

Pulsar rotation parameters may have obvious changes over long time. They are supposed to be measured using the latest timing data. Inaccurate astrometric parameters cause timing residual fluctuations that influence measurements of rotation parameters. Fluctuations caused by inaccurate timing astrometric parameters can be properly subtracted with independent and accurate astrometric parameters measured by the VLBI, leading to more accurate measurements of rotation parameters. Accurate proper motion and distance obtained with the VLBI also refine measurement of pulsar transverse velocity, which further leads to better estimates of influence from the Shklovsky effect on $\dot{P}$ and $\tau_{\rm c}$. $\tau_{\rm c}$ may obviously change after subtracting the Shklovsky effect for some pulsars, like PSR B0809$+$74 in our sample. Its increment is 2.5~Myr, which amounts to about 2\% of the $\tau_{\rm m}$. For some pulsars, like PSR B1540$-$06, increments of $\tau_{\rm c}$ are larger than differences between newly measured $\tau_{\rm m}$ and previously measured $\tau_{\rm p}$. With VLBI astrometric parameter solutions, glitch parameters can also be more accurately measured.

\begin{ack}
We would like to express our appreciation to Professor R.N. Manchester and Dr. M. Sekido for their good suggestions on this work. This work was supported in part by the National Natural Science Foundation of China (Grant Nos. U1631122, 11633007 and 11403073), the Strategic Priority Research Program of the Chinese Academy of Sciences (XDB23010200), the National Key R$\&$D Program of China (2018YFA0404602) and the Knowledge Innovation Program of the Chinese Academy of Sciences (KJCX1-YW-18). The hard work of all members of the TMRT team is vital for the high$\textrm{-}$quality observational data used in this paper.

\end{ack}

\end{document}